\documentclass[%
 reprint,
 amsmath,amssymb,
 aps]{revtex4-1}

\usepackage{graphicx}
\usepackage{dcolumn}
\usepackage{bm}

\newcommand{\beginappendix}{%
        \setcounter{table}{0}
        \renewcommand{\thetable}{A.\arabic{table}}%
        \setcounter{figure}{0}
        \renewcommand{\thefigure}{A.\arabic{figure}}%
            \setcounter{equation}{0}
        \renewcommand{\theequation}{A.\arabic{equation}}%
     }

\newcommand{\pa}[1]{\left(#1\right)}      
\newcommand{\abs}[1]{\left|#1\right|}

\newcommand{\Ex}[1]{\left\langle#1\right\rangle}

\newcommand{\e}{{\rm e}}

\newcommand{\com}[1]{}

\newcommand{\touch}{\emph{touch}~}
\newcommand{\Other}{\emph{other}~}

\newcommand{\F}{\phantom{.}_1F_1}

\begin{document}

\title{A generalized priority-based model for smartphone screen touches}




\author{Jean-Pascal Pfister}
\email{pfister@pyl.unibe.ch, corresponding author and Lead Contact}
\author{Arko Ghosh}%
 \email{a.ghosh@fsw.leidenuniv.nl, corresponding author}
\affiliation{%
\\
}%


\author{Jean-Pascal Pfister}
\affiliation{
Institute of Neuroinformatics and Neuroscience Center Zurich, University of Zurich / ETH Zurich\\
Department of Physiology, University of Bern} 

%
\author{Arko Ghosh}
\affiliation{%
Cognitive Psychology Unit, Institute of Psychology, Leiden University
}%


\date{\today}


\begin{abstract}
The distribution of intervals between human actions such as email posts or keyboard strokes demonstrates distinct properties at short vs long time scales. For instance, at long time scales, which are presumably controlled by complex process such as planning and decision making, it has been shown that those inter-event intervals follow a scale-invariant (or power-law) distribution. In contrast, at shorter time-scales - which are governed by different process such as sensorimotor skill - they do not follow the same distribution and little do we know how they relate to the scale-invariant pattern. Here, we analyzed 9 millions intervals between smartphone screen touches of 84 individuals which span several orders of magnitudes (from milliseconds to hours). To capture these intervals, we extend a priority-based generative model to smartphone touching events. At short-time scale, the model is governed by refractory effects, while at longer time scales, the inter-touch intervals are  governed by the priority difference between smartphone tasks and other tasks. The flexibility of the model allows to capture inter-individual variations at short and long time scales while its tractability enables efficient model fitting. According to our model, each individual has a specific power-low exponent which is tightly related to the effective refractory time constant suggesting that motor processes which influence the fast actions are related to the higher cognitive processes governing the longer inter-event intervals.
\end{abstract}

\maketitle







\section{Introduction}

Human actions such as mail correspondences, library loans or website visits are not equally distributed in time but are typically structured in bursts followed by long periods of inactivity \cite{Oliveira05a,Vazquez06a,Karsai18a}. 
Several types of models have been proposed to capture the power-law structure of inter-event time distribution (for a review see \cite{Karsai18a}). Priority-based queuing models \cite{Barabasi05a,Grinstein06a,Grinstein08a,Oliveira09a,Masuda09a,Min09a} rely on one (or multiple) list(s) of tasks to be executed, where each task is associated with a priority level which directly influences the timing of its execution. This class of models have been pioneered by Barabasi \cite{Barabasi05a} and then generalized to multiple interacting queues \cite{Oliveira09a,Min09a}, time-varying priorities \cite{Mryglod11a} or priorities which depend on the position within the list of tasks \cite{Vajna13a}. Those models provide an interesting interpretation for the origin of the power-law scaling for long intervals (they come from prioritizing tasks), but are usually not designed to capture short inter-event timings. 

On the other hand, Poisson-based models belong to another class \cite{Hidalgo-R06a,Vazquez07a,Malmgren08a,Guo11a}. They rely on the assumption that the event rate is governed by a Poisson process whose rate can change over time. Because of this dynamic rate assumption, those Poisson-based models can easily accommodate a precise description of short inter-event intervals. In the simplest case where the Poisson rates are piecewise constant and stochastically jump at each event time, the power-law exponent can be directly obtained from the distribution of Poisson rates \cite{Hidalgo-R06a}. This approach has been extended to continuously changing  Poisson rates \cite{Hidalgo-R06a,Vazquez07a,Malmgren08a,Guo11a}. In particular, the framework proposed by  \cite{Malmgren08a} provides a circadian explanation for the origin of power-law distributions. Self-exciting point processes (also called Hawkes processes \cite{Hawkes71a,Gilson18a}) have also been used to provide a mechanistic interpretation of power law inter-event intervals \cite{Masuda13a,Jo12a}. Those models are very flexible (they can accommodate both short and long inter-event intervals), but lack the priority-related interpretation. Indeed, it is unclear how those Poisson-based models relate to priority-based models.


Here, we start from a priority-based framework, generalise it on different levels and apply it to smartphone touchscreen interaction data (see Fig.~\ref{fig:data}). First, our model is the continuous-time extension of the classical priority-based model. Under this limit, we can compute analytically the inter-event distribution and show that our generalized priority-based model can be mapped to Poisson-based models. Secondly, our priority-based model does not only describe long inter-event intervals but also explicitly includes a relative refractory period after each event which helps to better describe short intervals and thereby overcomes the need to define an artificial onset of the power-law distribution \cite{Clauset09a}. Finally, because our model is based on arbitrary priority distribution and not on specific priority distribution imposed by the presence of lists (with discrete number of items), it can produce any power-law exponent. We found that for each subject, the inter-touch interval (ITI) distribution is different and well captured by the model. We also found that from those fitted parameters, we can quantify the relative priority placed on smartphone actions. \\

\begin{figure}[htbp]
\begin{center}
\begin{tabular}{ll}
{\bf a} & {\bf b} \\
\includegraphics[width=0.2\textwidth]{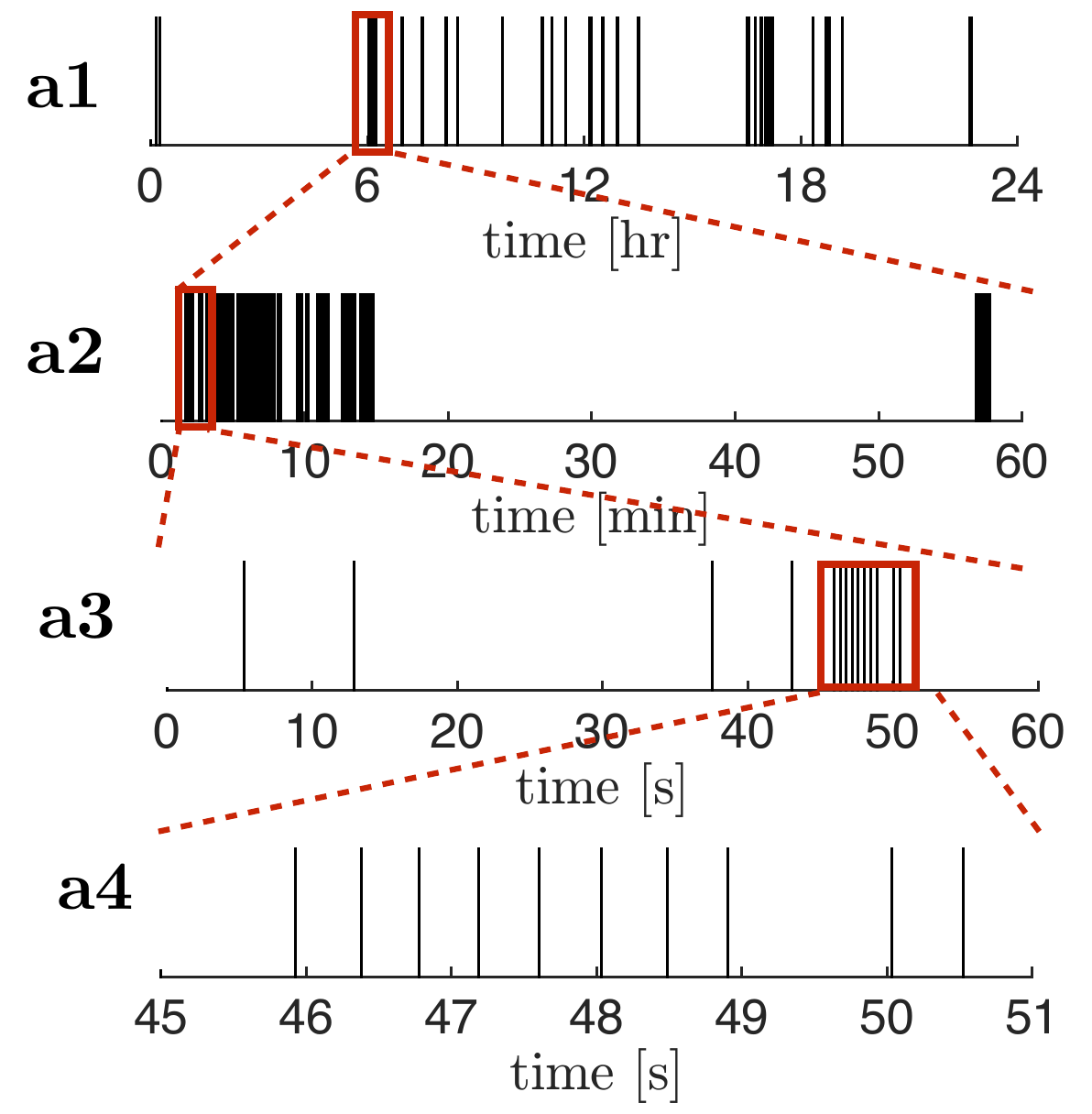}&
\includegraphics[width=0.25\textwidth]{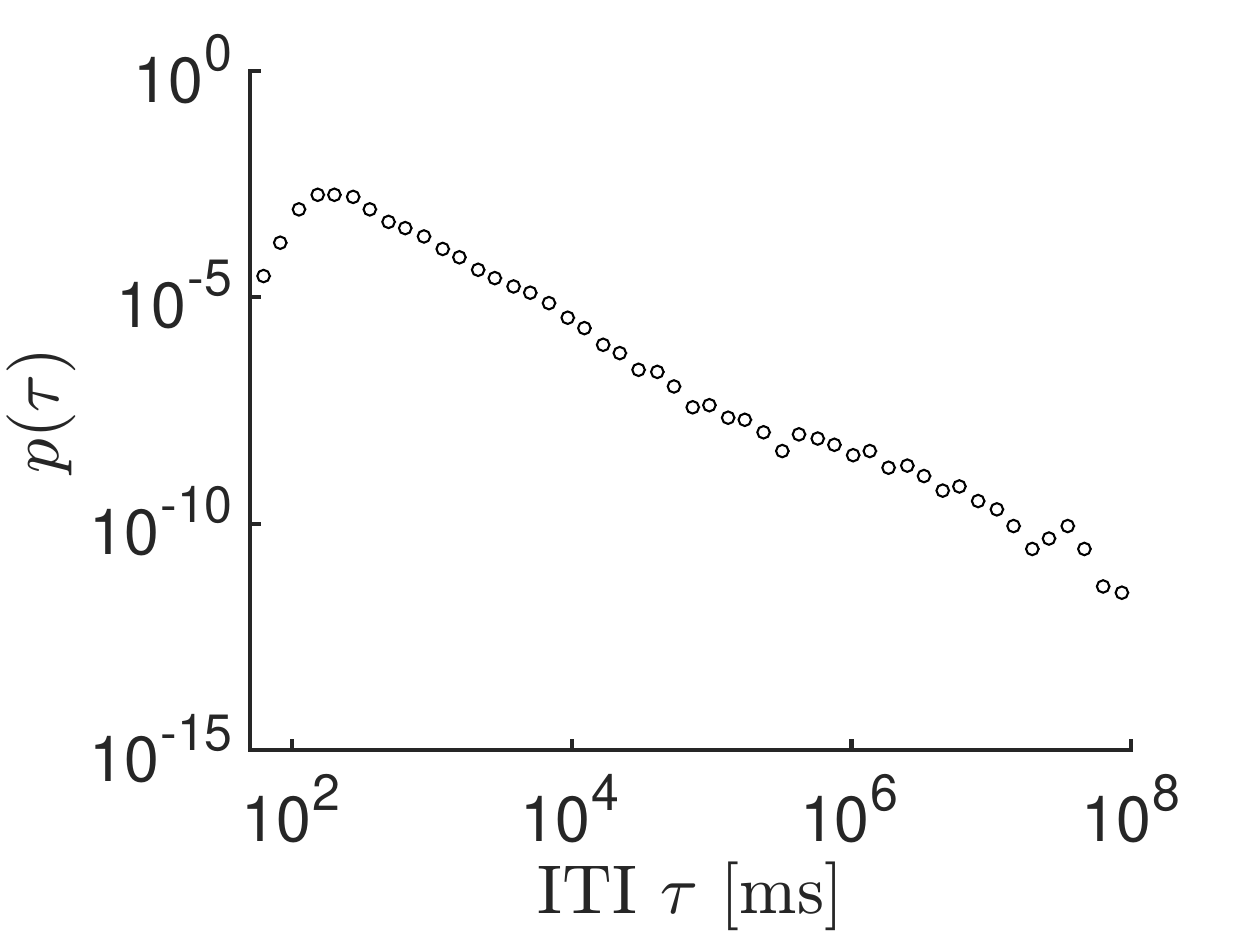}\\
\end{tabular}
\caption{\label{fig:data} Smartphone touch data. {\bf a} Smartphone touch events (vertical bars) are characterized by bursts as well as long gaps at time scales of hours ({\bf a1}) minutes ({\bf a2}) and seconds ({\bf a3}). At the sub-second time scale ({\bf a4}), touches are more regular. {\bf b} The inter-touch interval (ITI) distribution is scale free from seconds to hours. Data from one individual. }
\label{fig:data}
\end{center}
\end{figure}



\section{Smartphone Touching Model}
\subsection{Discrete-time model}
In the first step, we propose a discrete-time generative model for smartphone touches. This model extends existing priority-based models by including refractoriness \cite{Barabasi05a,Oliveira09a}.  The output of the model is the set of touch times $\{t_0,t_1,\dots,t_N\}$  where $t_i$ can take discrete values, i.e. $t_i = k_i\Delta t$ with $\Delta t$ being the bin width and $k_i\in \mathbb{N}$. Equivalently, the model output can be described by the touch train $s_t$
where $s_t = 1$ denotes the presence of a touch while $s_t = 0$ indicates the absence of a touch. 

Every touch is the result of a decision process. We assume that an individual can perform tasks from only two categories: either a task related to a smartphone screen \touch or \Other task such as driving a car. In each category, there can be important tasks (such as dialing an emergency number) or less important tasks (such as checking the news). So we will assume that every task can be described by its priority level which is a number between 0 and 1. Let $x_t\in [0,1]$ denote the priority associated with a \touch task at time $t$ and $y_t\in[0,1]$ the priority associated to the \Other task. If at time $t$ the \touch task associated to priority $x_{t}$ is executed (i.e. $s_{t} = 1$), then a new \touch task is considered and will be attributed a new \touch priority value drawn from the \touch priority distribution, i.e. $x_{\rm new}\sim p(x)$. If the \touch task is not executed ($s_{t} = 0$), its priority remains the same. This can be summarized as
\begin{equation}
x_{t+\Delta t} = x_t(1-s_t) + x_{\rm new}s_t \qquad x_{\rm new}\sim p(x). \label{eq:dx}
\end{equation}
Conversely, the dynamics for the \Other priority $y_t$ is such that when the screen is not touched at time $t$ (i.e. $s_t = 0$), then it is the \Other action that is executed and a new priority $y_{\rm new}$ must be drawn from $q(y)$. This is summarized as
\begin{equation}
y_{t+\Delta t} = y_ts_t + y_{\rm new}(1-s_t) \qquad y_{\rm new}\sim q(x).  \label{eq:dy}
\end{equation}
%
In order to generate a smartphone touch, two conditions need to be satisfied. Firstly the priority $x_t$ of the smartphone action needs to be greater than the priority $y_t$ of the other action and secondly, the individual must be in a non-refractory state. Formally, the touch variable $s_t$ is sampled from the following Bernoulli distribution:
\begin{equation}
s_t \sim {\rm Bernoulli}(\lambda(x_t,y_t,\tau_t)\Delta t), \label{eq:s}
\end{equation}
where the touching intensity $\lambda$ (probability per time bin $\Delta t$) is given by
\begin{equation}
\lambda(x,y,\tau) = \rho r(\tau) H(x-y), \label{eq:lambda}
\end{equation}
where $\tau = t-\hat{t}$ is the time since last touch ($\hat{t} = \max_{t_k}\{t_k<t\}$) and $H$ is the Heaviside step function which guarantees that touches can only be generated when $x>y$ and $\rho$ is the touching rate. $r(\tau)\geq 0$ is the refractory function which includes post-touch effects (i.e. right after a touch, the touch probability can be reduced). A hard refractoriness function takes the following form
\begin{equation}
r(\tau) = H(\tau-\Delta)
\end{equation}
where $H(\tau)$ is the Heaviside step function (i.e. $H(\tau) = 1$ if $\tau\geq 0$ and $H(\tau)$ else) and $\Delta$ the hard refractory time (i.e. minimal ITI). If we relax this strong condition and allow touches for any $\tau>0$ (but with reduced probability when $\tau\simeq 0$), we can define a relative refractoriness function as a sum of basis functions:
\begin{equation}
r(\tau) = 1 + \sum_{k=1}^n\gamma_k\exp\pa{-\alpha_k\tau},
\end{equation}
with logarithmically spaced inverse time constants, i.e. $\alpha_k = \alpha_1\beta^{-(k-1)}$. We took $\alpha_1^{-1} = 50$ ms, and set $\beta$ such that $\alpha_n^{-1} = 1000$ ms. Note that the set $\{\gamma_k\}_{k=1}^n$ has to be chosen such that the condition $r(\tau)\geq 0$ is satisfied for all $\tau\geq 0$. If $r(0)\leq 0.5$, we define the effective time constant $\tau^*$ as the time for which refractoriness is half, i.e. $r(\tau^*) = 0.5$; see Fig.~(\ref{fig:results}b1). In the rare cases where multiple solutions exists for $\tau^*$ (which can occur when $r(\tau)$ is non-monotonic), we took the maximal value of the set of solutions. 

The discrete-time model described by Eqs~(\ref{eq:dx}), (\ref{eq:dy}) and (\ref{eq:s}) is a latent dynamical system. 
Note that sampling this model is slow since the complexity of this sampling scheme scales with the number of bins. Even more critical is the learning procedure for such a latent dynamical model which can be prohibitively slow for smartphone touching data sets which typically extend over months. A much faster sampling scheme is proposed below.

\subsection{Continuous-time model}
The idea of the continuous-time model is to directly sample the intervals $\tau$ instead of sampling the touch variable  $s_t$ at each time step. The transition to this continuous model can be done in two steps. First, we observe that when $\Delta t$ is small, the \Other priorities $y_t$ constantly change (except at the rare times where $s_t = 1$), i.e. Eq.~(\ref{eq:dy}) can be approximated as $y_t\sim q(y)$. 
%
%
This means that the priorities $y_t$ are independent of time and therefore, the probability of generating a touch can be marginalized over $y_t$:
\begin{eqnarray}
p(s_t|x_t,\tau_t) &=& \int_0^1p(s_t|x_t,y_t,\tau_t)q(y_t)dy_t \nonumber \\
&=& {\rm Bernoulli}(\bar{\lambda}(x_t,\tau_t)\Delta t),
\end{eqnarray}
where the average touching intensity $\bar{\lambda}$ is given by
\begin{equation}
\bar{\lambda}(x,\tau) = \int_0^1\lambda(x,y,\tau)q(y)dy  = \rho r(\tau) \pi(x),
\end{equation}
and $\pi(x)$ is the probability of having $x>y$ for a given $x$
\begin{equation}
\pi(x) = \int_0^x q(y)dy.
\end{equation}
In the second step, we take the limit $\Delta t\rightarrow 0$ and therefore, the inter-touch interval distribution conditioned on $x$ can be expressed as (see also \cite{Bremaud81a}):
\begin{equation}
p(\tau|x) = \bar{\lambda}(x,\tau)\exp\pa{-\int_0^\tau\bar{\lambda}(x,t)dt} \label{eq:ptaux}.
\end{equation}
The unconditioned ITI distribution is obtained by averaging the conditioned ITI distribution over the \touch priority distribution $p(x)$:

\begin{equation}
p(\tau) = \int_0^1 \bar{\lambda}(x,\tau)\exp\pa{-\int_0^\tau\bar{\lambda}(x,t)dt}p(x)dx. \label{eq:ptau}
\end{equation}

So samples of the continuous-time model can be simply obtained in a two-step procedure. First, $x$ is sampled from $p(x)$, then $\tau$ is sampled from $p(\tau|x)$ given by Eq.~(\ref{eq:ptaux}). For this second step, one can use the time rescaling theorem \cite{Bremaud81a}. Note that this continuous-time model describes a renewal process and hence the sampling complexity scales with the number of touches $N$.

In the absence of refractoriness (i.e. $r(\tau) = 1$), the sampling procedure is even simpler. First, a Poisson rate $\bar{\lambda} = \rho x$ can be drawn from a distribution of rates $p(\bar{\lambda})$ with maximal rate $\bar{\lambda}_{\rm max} = \rho$, then $\tau$ is sampled from an exponential distribution $p(\tau |\bar{\lambda}) = \bar{\lambda}\exp(-\bar{\lambda}\tau)$ and the ITI can be expressed as  

\begin{equation}
p(\tau) = \int_0^\rho p(\tau|\bar{\lambda})p(\bar{\lambda})d\bar{\lambda}
\end{equation}

which is precisely the ITI one would get from an heterogeneous Poisson model \cite{Hidalgo-R06a}.  This shows the equivalence between the priority-based model and the Poisson-based models. 

\section{Properties of the model}
\subsection{Invariance of the model}
 Before giving a parametric form for all distributions, let us first note an invariant property of the model. In particular, it can be shown (see appendix \ref{sec:inv}) that the ITI distribution given by Eq.~(\ref{eq:ptau}) remains unchanged if the pair of priority distributions $(p(x),q(y))$ is replaced by $(\tilde{p}(x),\tilde{q}(y))$ given by
\begin{equation}
\tilde{p}(x) = p(\phi(x))\phi'(x)\quad{\rm and}\quad \tilde{q}(y) = q(\phi(y))\phi'(y) \label{eq:ptilde}
\end{equation}
where $\phi$ is a differentiable and strictly monotonously increasing function with boundary conditions $\phi(0)=0$ and $\phi(1)=1$. This invariance can be understood intuitively by noting that the notion of priority contains some arbitrariness. Indeed, the only element which is relevant in the decision process is whether $x$ is larger or smaller than $y$ (see Eq.~(\ref{eq:lambda}). If we define a new priority $x' = \phi(x)$ (with the above conditions on $\phi$), we observe that the ordering remains unchanged, i.e. $x>y \Rightarrow \phi(x)>\phi(y)$.  This observation can also be made more formally with a change of variable in Eq.~(\ref{eq:ptau}) (see Appendix \ref{sec:inv}). %
%
%
%
Secondly, this invariance property of the model means that without loss of generality, we can set one distribution and rescale the other one. For example, without loss of generality, we can set $q(y) = 1$. For the \touch priority distribution, we will assume that it is given by a  Beta distribution:
\begin{equation}
p(x) = {\rm Beta}(x;a,b) = \frac{x^{a-1}(1-x)^{b-1}}{B(a,b)}
\end{equation}
where $B(a,b) = \int_0^1x^{a-1}(1-x)^{b-1}dx$ is the Beta function. With the above choice of $q$, the ITI distribution in Eq.~(\ref{eq:ptau}) can be rewritten in a simpler form
\begin{equation}
p(\tau) = \rho r(\tau) \int_0^1  x\exp\pa{-x\rho\int_0^\tau r(t) dt}p(x)dx \label{eq:ptau2}
\end{equation}

\subsection{Scale free inter-touch interval distribution}
For short time scales ($\tau<\alpha_n^{-1}$), the ITI distribution is governed by the refractory function $r$ (see Fig.~\ref{fig:model}a). However, for longer time scales ($\tau\gg\alpha_n^{-1}$), the ITI distribution follows a power-law distribution. This can be seen in two steps. First, in the limit of large $\tau$, we have $r(\tau)\rightarrow 1$. Secondly, in the limit of large $\tau$, we know from Eq.~(\ref{eq:ptau2}) that the ITI distribution is only sensitive to the \touch priority distribution in the vicinity of $x = 0$ that we denote as  $p_0(x)$. Note that $p_0(x)$ is not normalized. For the Beta distribution, we have $p(x)\rightarrow p_0(x) = x^{a-1}/{\rm B}(a,b)$ when $x\rightarrow 0$. Therefore, when $\tau\gg \alpha_n^{-1}$, the ITI distribution can be approximated as
\begin{eqnarray}
p(\tau) &\simeq& \rho\int_0^1xp_0(x)\e^{-x\rho\tau}dx \nonumber \\
&\simeq& \frac{\Gamma(a+1)}{{\rm B}(a,b)\rho^{a}}\tau^{-(a+1)},
\end{eqnarray}
where  $\Gamma(z) = \int_0^{\infty}x^{z-1}\e^{-x}dx$ is the Gamma function. Therefore the power-law exponent is given by  $a+1$ (see Fig.~\ref{fig:model}b). So when $a$ is large, \emph{touching} tasks are more important - in the sense that there are few low-priority touch tasks - and consequently there are less large ITI. 

\begin{figure}[htbp]
\begin{center}
\begin{tabular}{lll}
{\bf a} & {\bf b}\\
\includegraphics[width=0.23\textwidth]{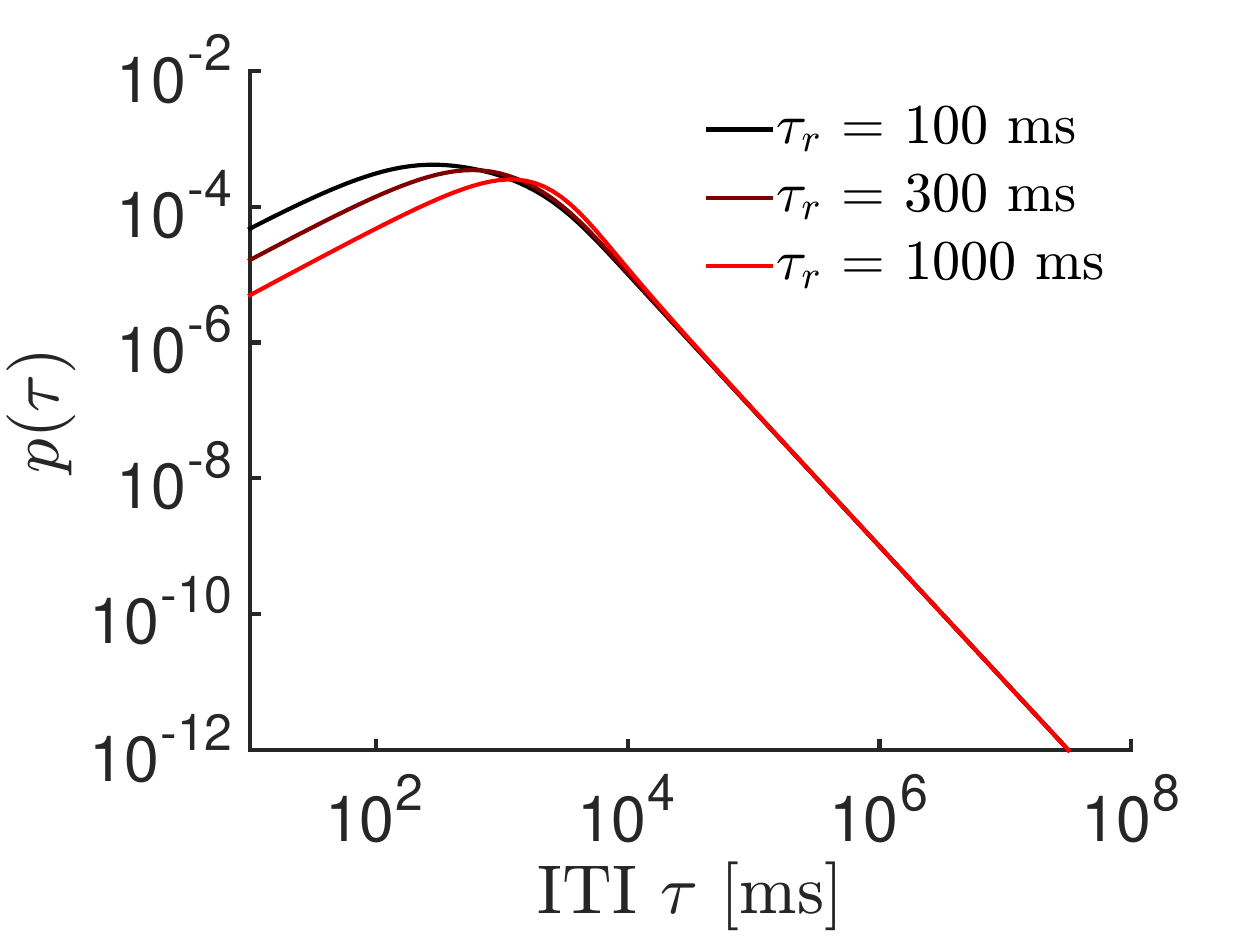} &
\includegraphics[width=0.23\textwidth]{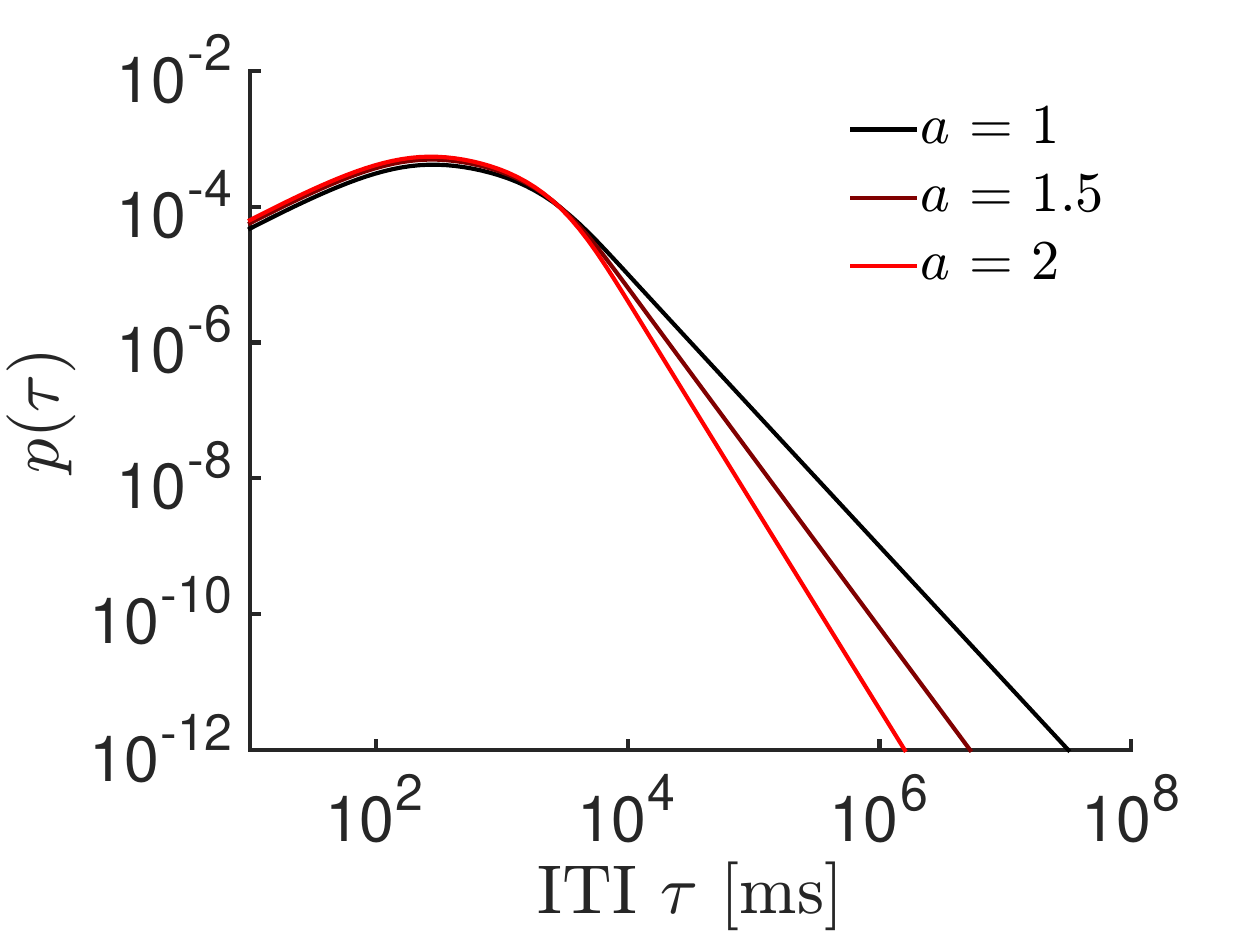}
\end{tabular}
\caption{\label{fig:model} Properties of the smartphone touching model.  {\bf a} The refractory time constant affects the early part of the ITI distribution. $n=1$, $\tau_r = \alpha_1^{-1}$.  {\bf b} The parameter $a$ from the priority distribution affects the power-law exponent of the ITI distribution.  }
\label{fig:model}
\end{center}
\end{figure}


\section{Results}
\subsection{Model fitting}
For each subject, we fitted models with hard refractoriness and relative refractoriness. For each of those model classes, we fitted several variants by fixing a subset of parameters so specific values (see Table~\ref{tab:models}). In particular, we fitted 6 different models:

\begin{enumerate}
\item Model $M_1$ is the simplest model and contains only 2 parameters: $\theta = (a,\rho)$. It is assumed that $b=1$ and that there is no refractoriness ($\Delta = 0$). 
\item  Model $M_2$ is the same as model 1 except that the \touch priority distribution has 2 free parameters: $a$ and $b$. Overall, it contains 3 parameters:  $\theta = (a,b,\rho)$.
\item Model $M_3$ includes hard refractoriness (with refractory time $\Delta$) but assumes $b=1$. It contains therefore 3 parameters:  $\theta = (a,\rho,\Delta)$.
\item Model $M_4$ is the same as model 3 except that the \touch priority distribution is described by both $a$ and $b$. It contains 4 parameters: $\theta = (a,b,\rho,\Delta)$.
\item Model $M_5$ uses a relative refractory kernel parametrized by $n$ basis functions with coefficients $\gamma_1,\dots,\gamma_n$. It also assumes that $b=1$. So the model contains $n+2$ parameters: $\theta = (a,\rho,\gamma_{1\dots n})$.
\item Model $M_6$ is the same as model $M_5$, but $b$ is not constrained to be equal to one. The models contains therefore $n+3$ parameters: $\theta = (a,b,\rho,\gamma_{1\dots n})$.
\end{enumerate}

\begin{table}[hb]
\begin{center}
\begin{tabular}{l | l | c | l}
Model  & parameters & \# of param. & assumptions\\ 
\hline
$M_1$ &  $\theta = (a,\rho)$ & 2 & $r$: hard, $b=1,\Delta=0$\\
$M_2$ &  $\theta = (a,b,\rho)$ & 3 & $r$: hard, $\Delta = 0$\\
$M_3$ &  $\theta = (a,\rho,\Delta)$ & 3 &$r$: hard,  $b=1$\\
$M_4$ &  $\theta = (a,b,\rho,\Delta)$ & 4 & $r$: hard\\
$M_5$  & $\theta = (a,\rho,\gamma_{1\dots n})$ & n+2 & $r$: rel, $b=1$\\
$M_6$ &  $\theta = (a,b,\rho,\gamma_{1 \dots n})$ & n+3 & $r$: rel
\end{tabular}
\end{center}
\caption{\label{tab:models} List of models. }
\end{table}



For each model and for each subject, the model parameters $\theta$ are fitted from the set  $\mathcal{D} = \{\tau_i\}_{i=1}^{N}$  of inter-touch intervals $\tau_i = t_{i}-t_{i-1}$. In order to do  so, we relied on the continuous-time model which massively simplifies the expression of the log-likelihood. Indeed, the detailed model can be seen as a dynamical latent variable model (where the latent variables are $x$ and $y$) which can be fitted through EM type algorithm but is known to be very slow. Here, because of the analytical expression of the ITI for the continuous-time model (see Eq.~\ref{eq:ptau2}), we can express the following objective function 
\begin{equation}
\mathcal{L}(\theta) = L(\theta) - \lambda\sum_{k=1}^n\gamma_k^2, \label{eq:logpost}
\end{equation}
which is the log-likelihood $L(\theta) = \sum_{i=1}^N\log p(\tau_i)$ (see Eq.~\ref{eq:LL})
minus a regularization term on the coefficients $\gamma_k$ to prevent overfitting. This regularization term (with $\lambda = 1000$) is only used in models 5 and 6. Note that this objective function can be seen as the log-posterior with a Gaussian prior (with variance $1/2\lambda$) on the coefficients $\gamma_k$ and a flat prior for the other parameters. 

Because the refractory kernel must remain positive for all time, i.t. $r(\tau)\geq 0$, $\forall \tau\geq 0$, the optimization task can be expressed as
\begin{eqnarray}
\theta^* &=& \arg\max_{\theta} \mathcal{L}(\theta)\nonumber \\
 &{\rm s.t.}&  \quad \sum_{k=1}^n\exp(-\alpha_k\tau)\gamma _k^*\geq -1 \quad \forall \tau\geq 0, \label{eq:optim}
\end{eqnarray}
However, the difficulty of the optimization problem defined in Eq.~(\ref{eq:optim}) lies in the fact that the constraints are defined for all $\tau\geq 0$ (i.e. infinitely many inequality constraints). For a practical numerical implementation, we defined a grid of $M=200$ points $\tau_1,\dots,\tau_M$ where the first 100 points are linearly spaced ($\tau_i = i\Delta t$ for $1\leq i< 100$ and $\Delta t = 1$ ms) and the subsequent 100 points are logarithmically spaced ($\tau_i = \kappa^{i-100}\tau_{100}$ for $100 < i \leq 200$ where $\kappa$ is set such that $\tau_{200} = 3\alpha_{n}^{-1}$). So we replace the inequality constraints of Eq.~(\ref{eq:optim}) by
\begin{eqnarray}
\sum_{k=1}^n\exp(-\alpha_k\tau_i)\gamma _k^*\geq -1 \quad \forall i=1,\dots,M   \nonumber
\end{eqnarray}

\subsection{Fitting results}
We recorded smartphone touches from 84 individuals for an average duration of 36.5 days (see Appendix \ref{sec:data} for details on data collection). The average number of smartphone screen touches per day ranged from 285 to 9'915 with a median value of 2'540 touches per day. 

\begin{figure}[htbp]
\begin{center}
\begin{tabular}{lll}
{\bf a} & {\bf b}\\
\includegraphics[width=0.23\textwidth]{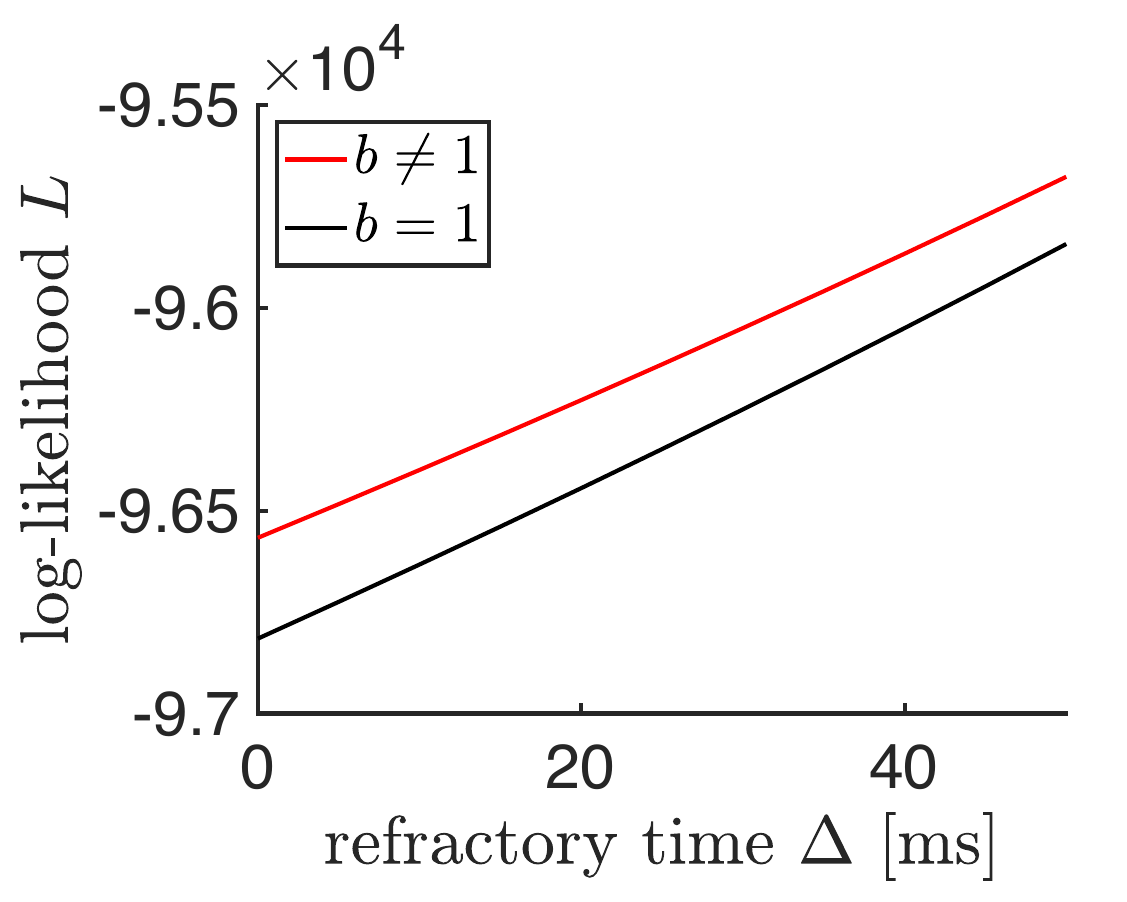} &
\includegraphics[width=0.23\textwidth]{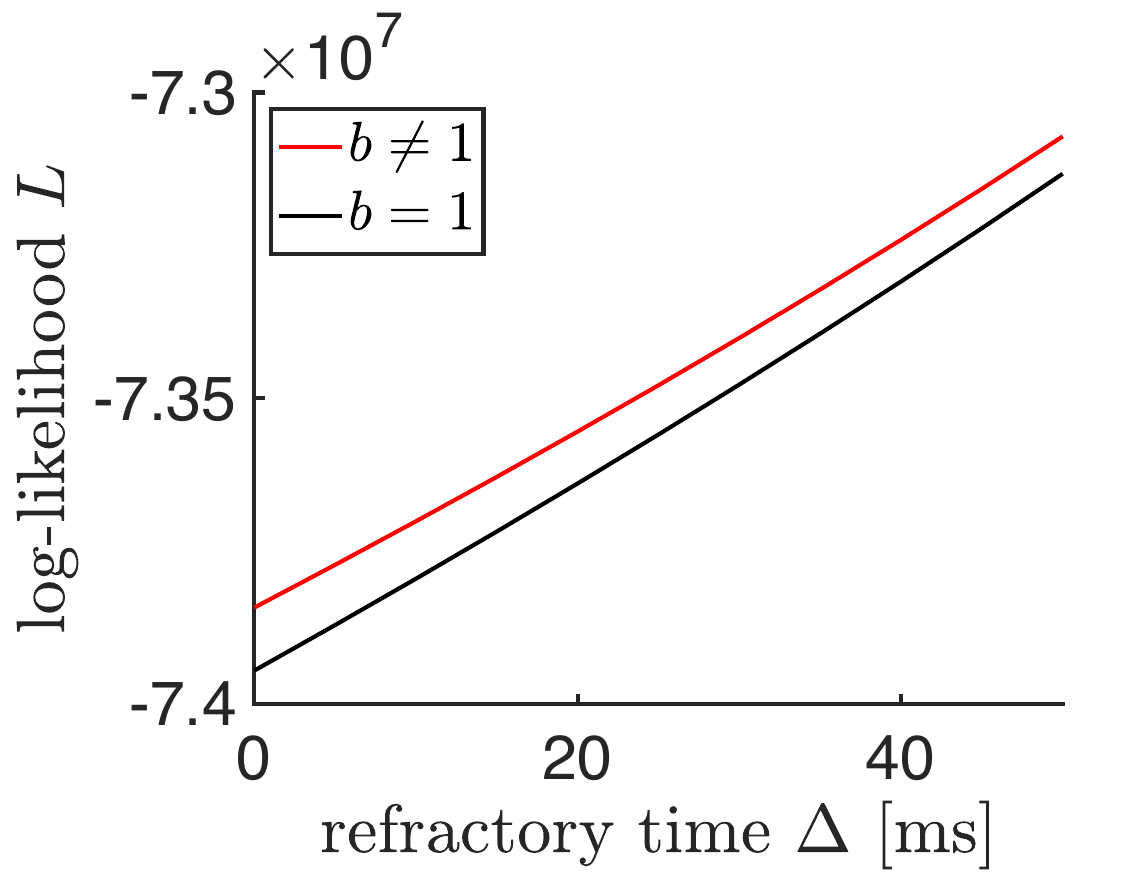}\\
{\bf c} & {\bf d}\\
\includegraphics[width=0.23\textwidth]{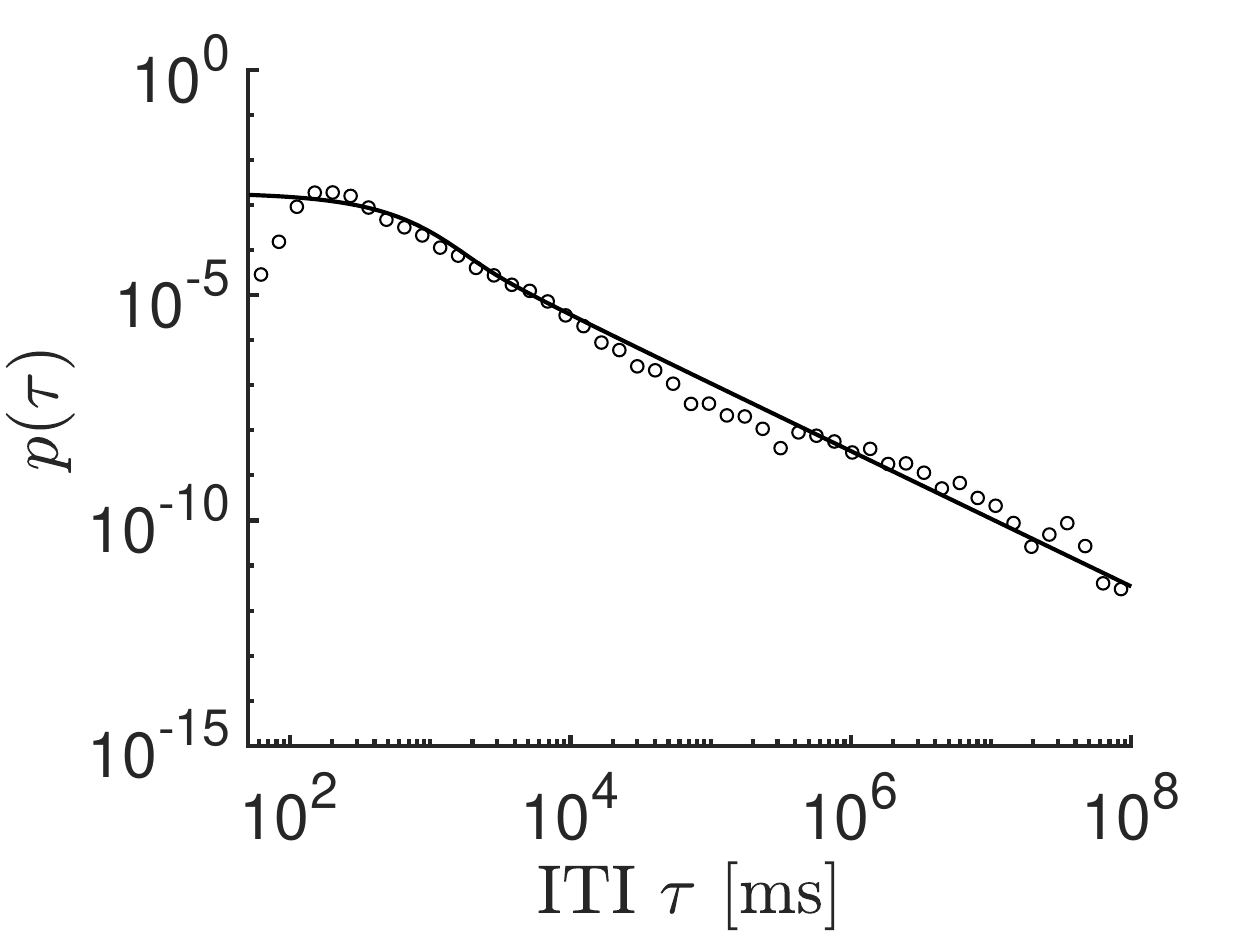} &
\includegraphics[width=0.23\textwidth]{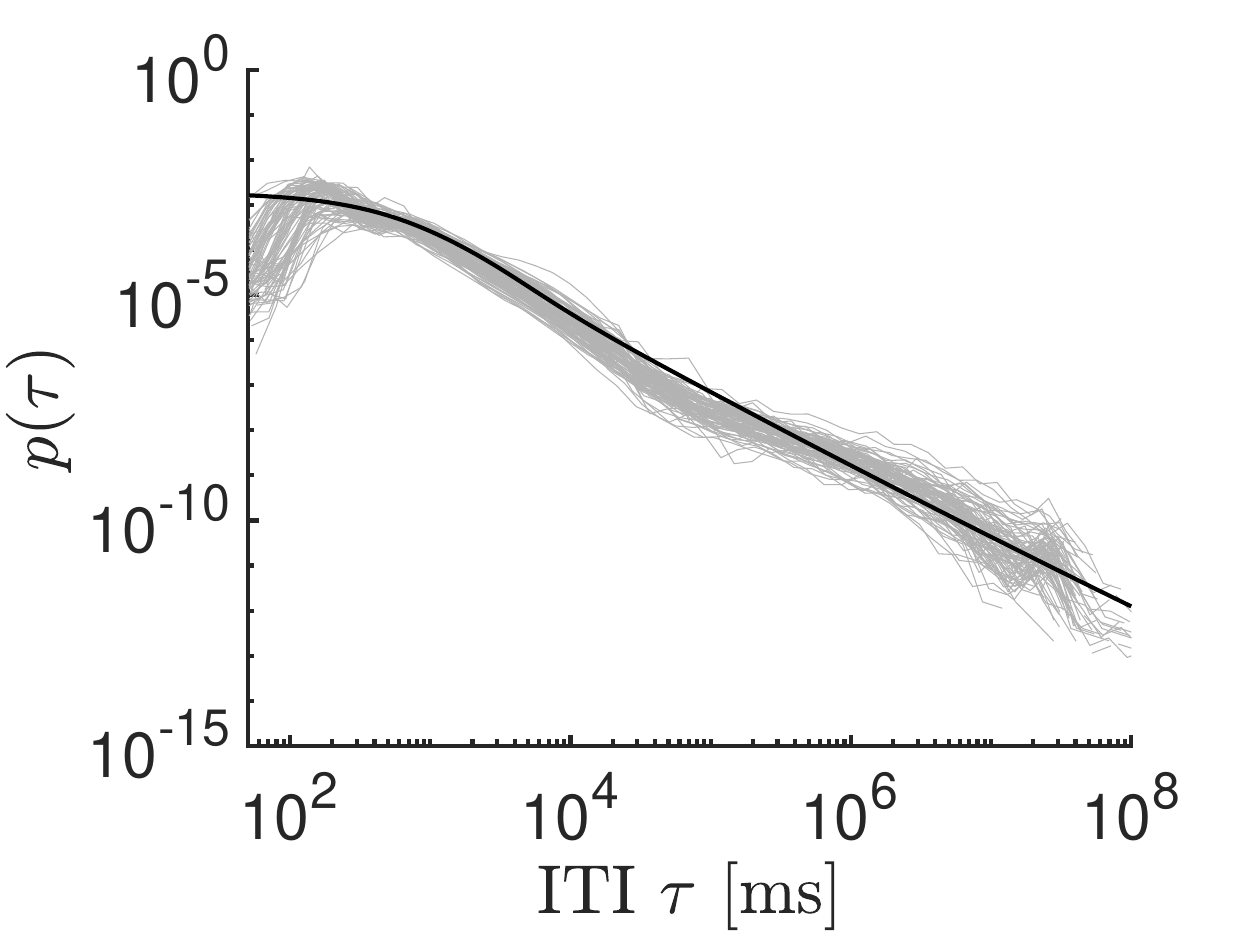}
\end{tabular}
\caption{\label{fig:hard} Fitting results for the models with hard refractoriness ($M_3$ and $M_4$). {\bf a}. Log-likelihood of model $M_3$ (black) and $M_4$ (red) as a function of the refractory time $\Delta$ for a single subject.  The best refractory time is $\Delta^* = 50$ ms. {\bf b}. The log-likelihood summed across subjects also elicits an optimal refractory time at $\Delta = 50^*$ ms for both model $M_3$ (black) and $M_4$ (red). {\bf c}. Inter-tap interval distribution for one subject (solid line: fit, circles: data). {\bf d}. Inter-touch interval distribution across the whole population. Each gray line corresponds to the data from one subject. Solid black line denotes the median model ITI distribution.}
\end{center}
\end{figure}

For each individual, the 6 different models were fitted according to the procedure described above. In particular, we first fitted the models without refractoriness ($M_1$ and $M_2$) and the models with hard refractoriness ($M_3$ and $M_4$). We found that the likelihood can be drastically improved by adding the hard refractory time parameter $\Delta$ (see Fig.~\ref{fig:hard}a, b and Fig.~\ref{fig:s1}a). Actually, the optimal value is exactly $\Delta^* = \tau_{\rm min}$ where $ \tau_{\rm min}$ is the minimal ITI (see Eq.~\ref{eq:dLDelta}). The fitted ITI for model $M_4$ (see Fig.~\ref{fig:hard}c and d) is decent, but short ITI are not well captured. 

\begin{figure}[htbp]
\begin{center}
\begin{tabular}{lll}
{\bf a1} & {\bf a2}\\
\includegraphics[width=0.23\textwidth]{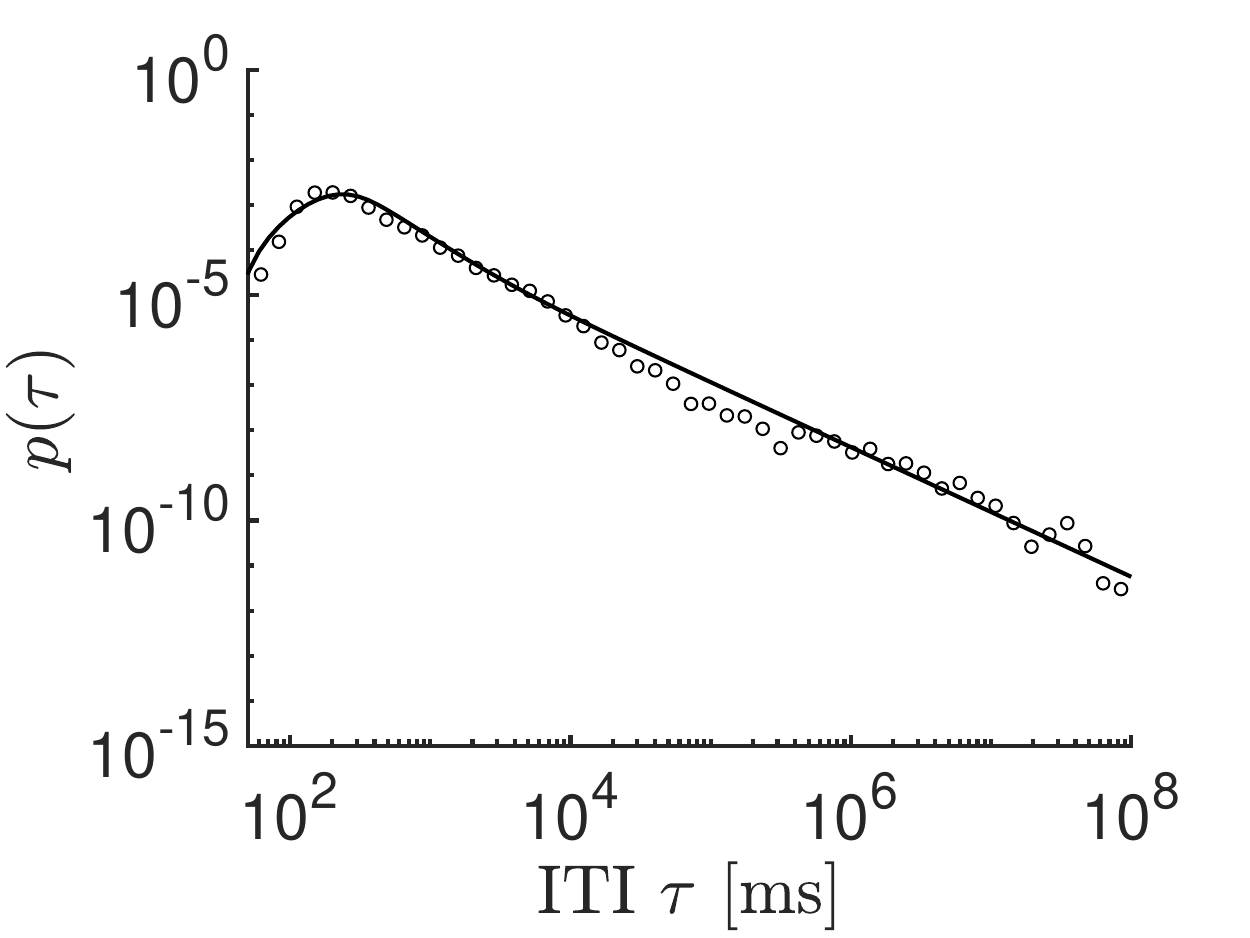}&
\includegraphics[width=0.23\textwidth]{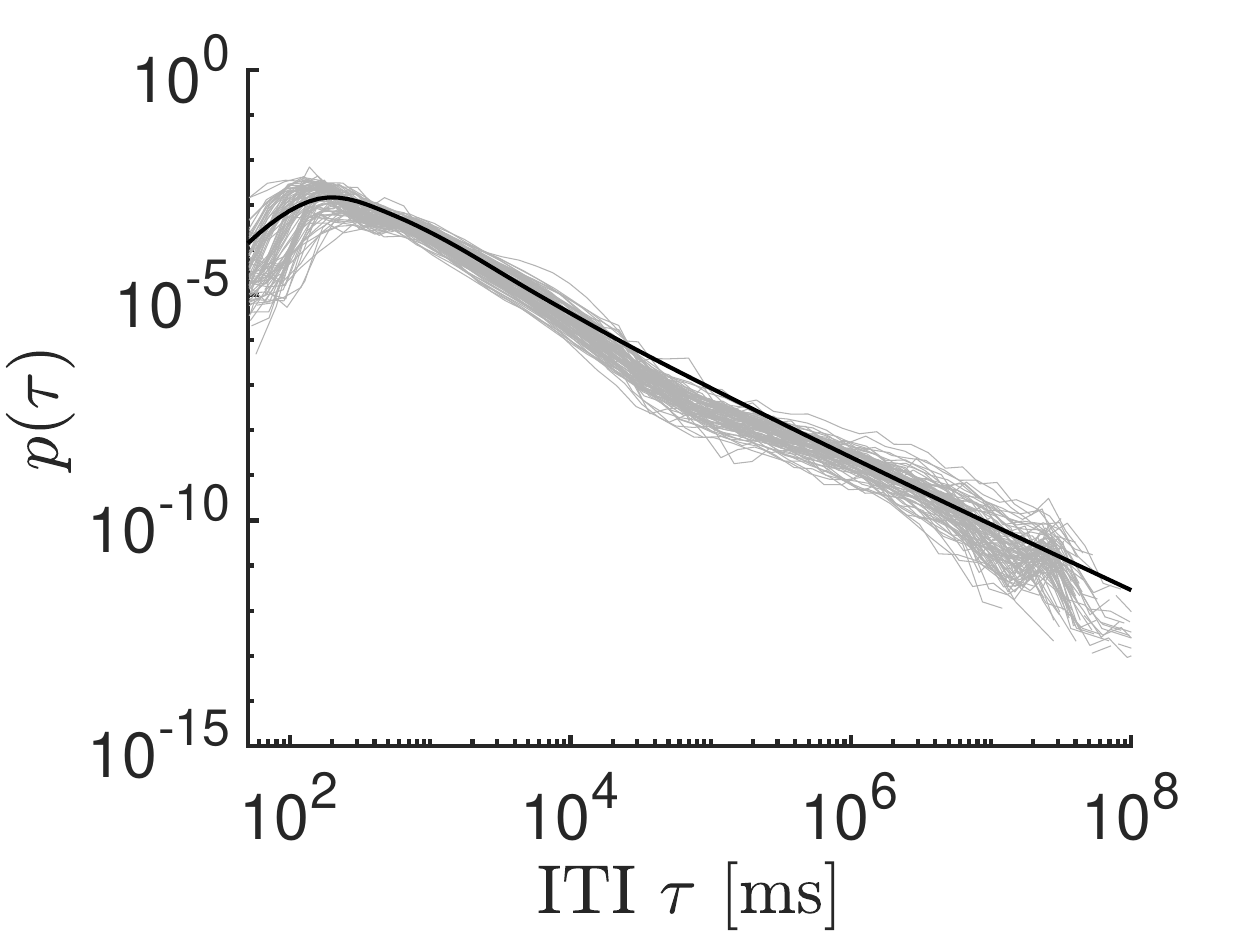}\\
{\bf b1} & {\bf b2}\\
\includegraphics[width=0.23\textwidth]{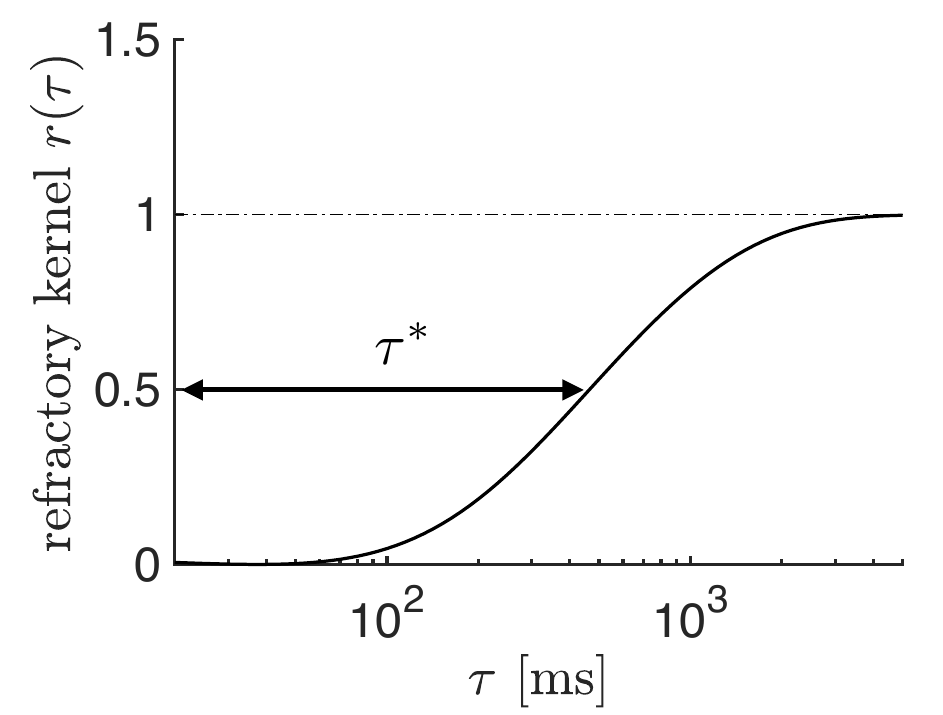} &
\includegraphics[width=0.23\textwidth]{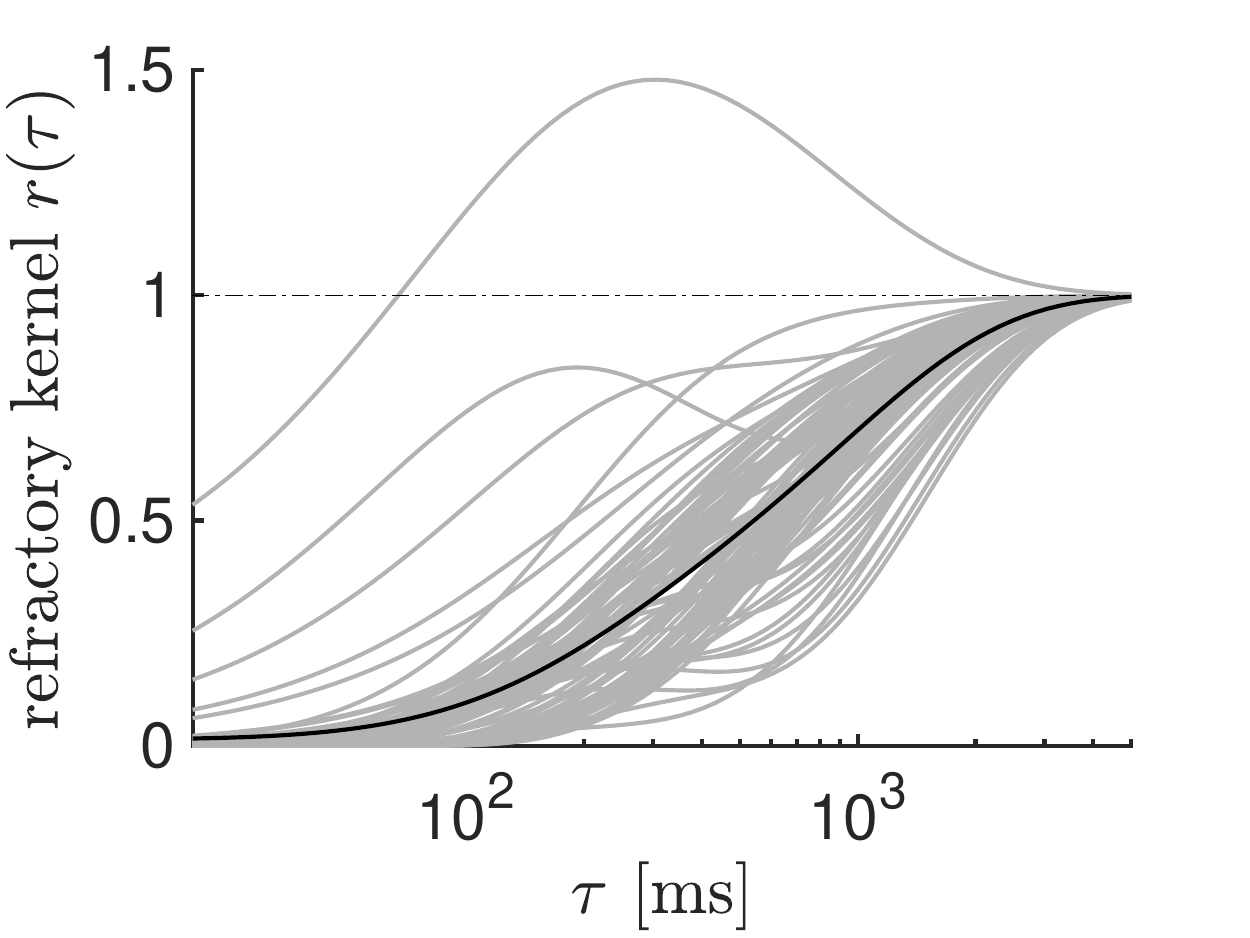} \\
{\bf c1} & {\bf c2}\\
\includegraphics[width=0.23\textwidth]{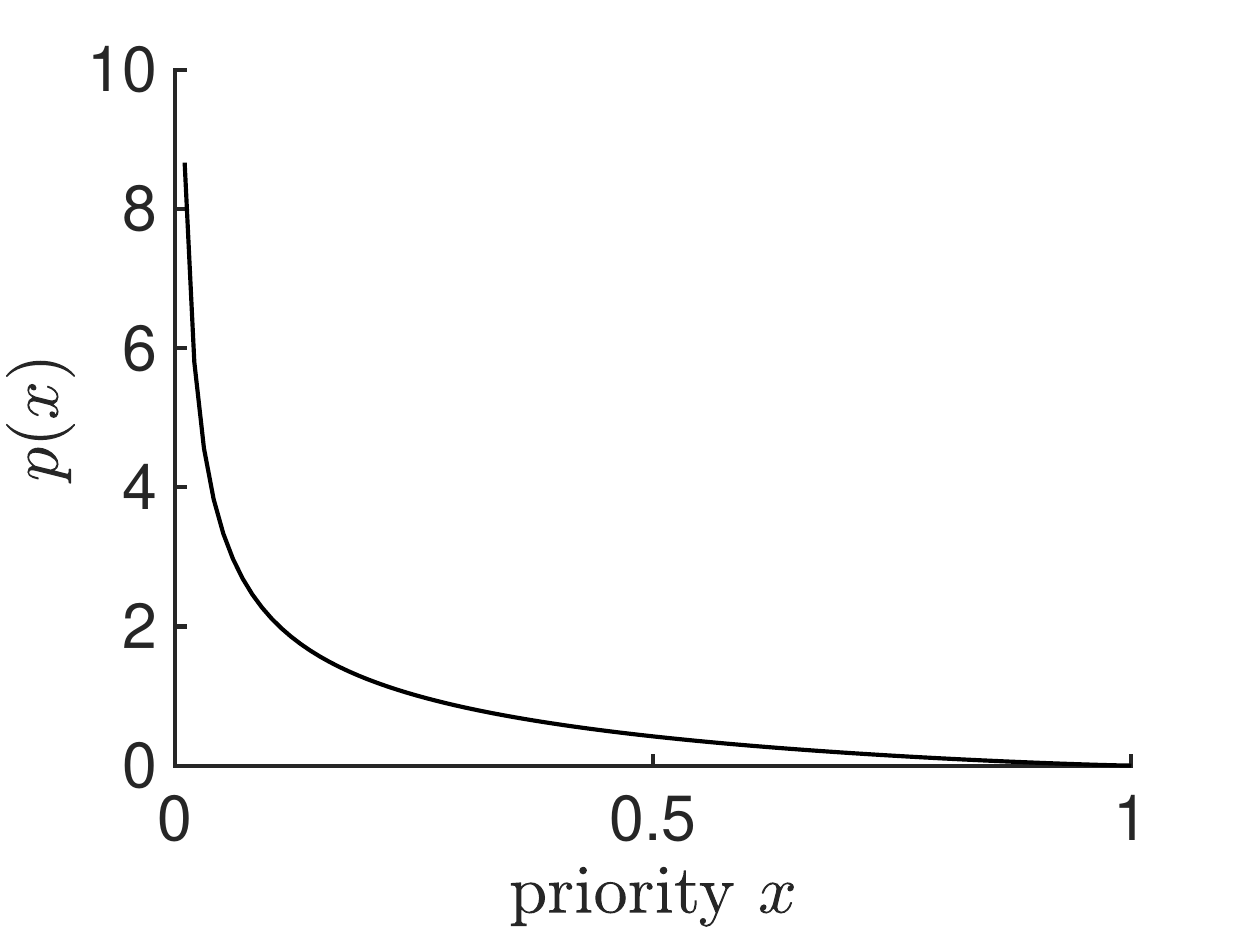} &
\includegraphics[width=0.23\textwidth]{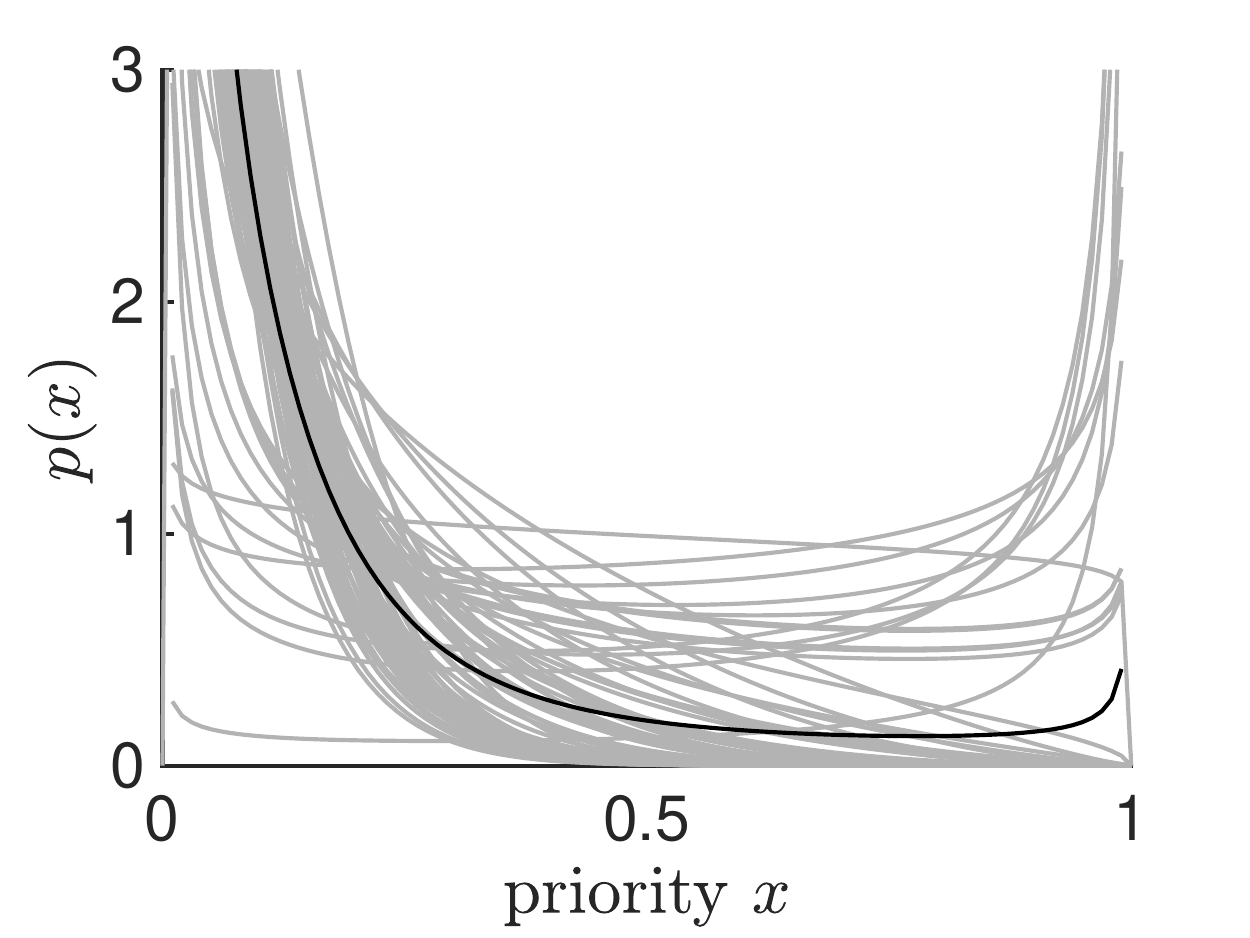} \\
{\bf d} & {\bf e}\\
\includegraphics[width=0.23\textwidth]{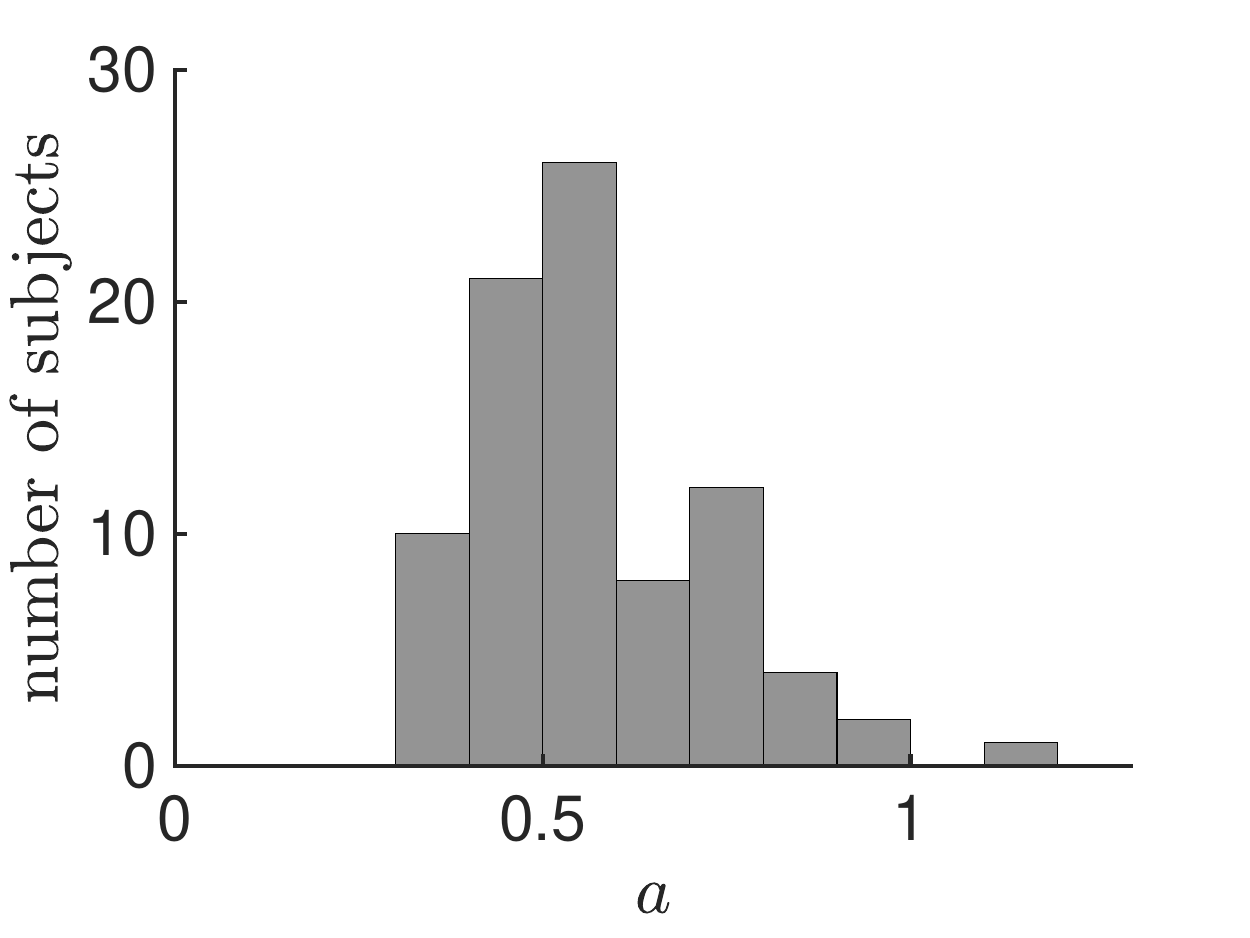}&
\includegraphics[width=0.23\textwidth]{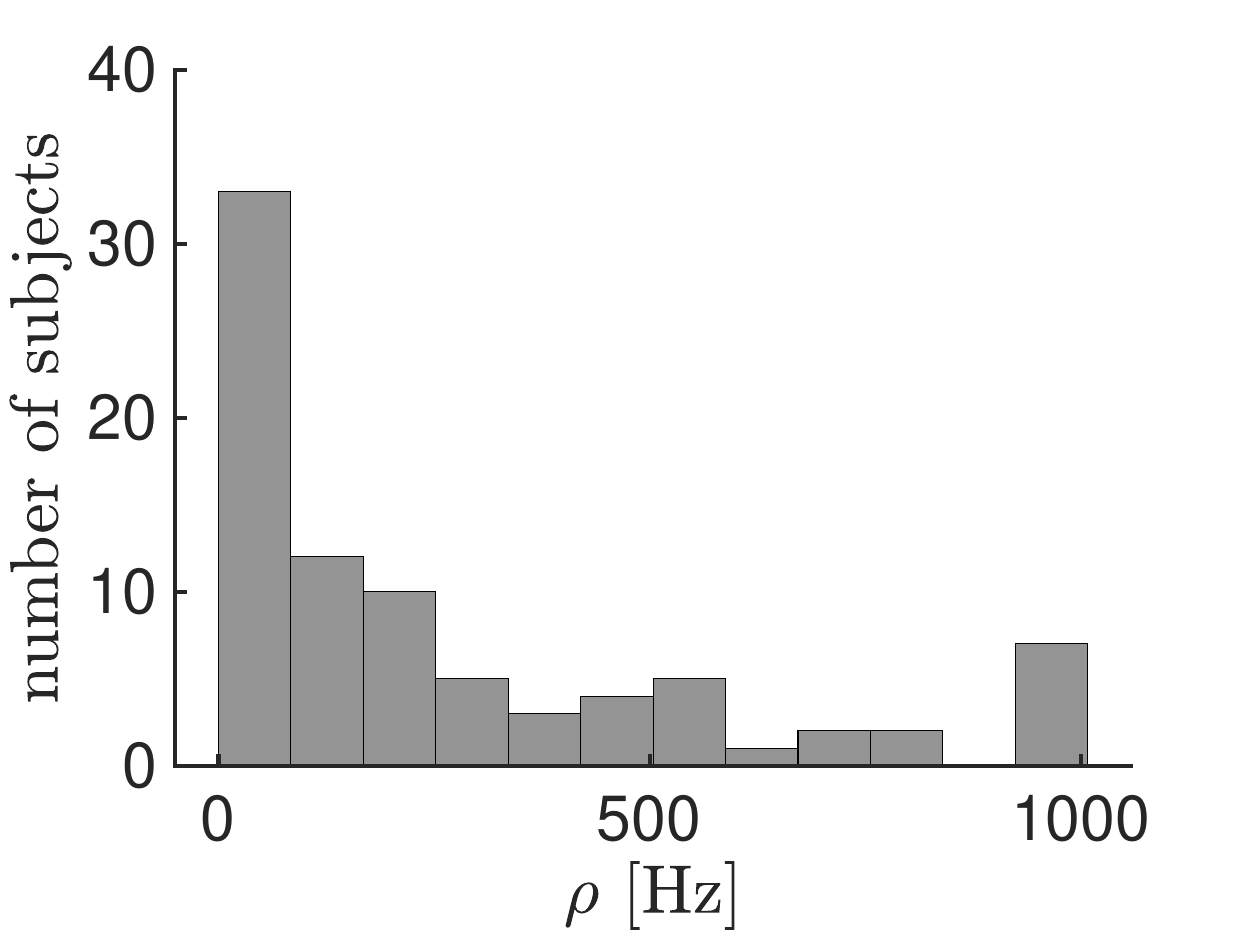}
\end{tabular}
\caption{Fitting results of model $M_6$ (with $n=21$) for one subject ({\bf a1}-{\bf c1}) and for the population of 84 subjects ({\bf a2}-{\bf c2}).  {\bf a1} The ITI distribution for one given subject (open circles) is well captured model (solid line). {\bf b1} Refractory kernel. The effective time constant $\tau^*$ is defined as $r(\tau^*) = 0.5$ {\bf c1}  \touch priority distribution (with $q=1$).
{\bf a2}-{\bf c2} Same as in {\bf a1}-{\bf c1} but for each of the 84 subjects (gray lines). Solid lines denote the median ITI ({\bf a2}), refractory kernel ({\bf b2}) and priority distribution ({\bf c2}).
{\bf d} Distribution of the parameter $a$. {\bf e} Distribution of the touching rate $\rho$. }

\label{fig:results}
\end{center}
\end{figure}

We then fitted the models with relative refractoriness ($M_5$ and $M_6$) and displayed the fitting results of the best model ($M_6$ with $n=21$ basis functions); see Fig~\ref{fig:results}.  We found that for each individual the empirical ITI distribution (see Fig.~\ref{fig:results}a1) is well captured by the model both for the short time scales (up to 1s) which is strongly influenced by the refractory kernel $r(t)$ (see  Fig.~\ref{fig:results}b1) as well as the longer ITI which has a typical power-law decay. Note that because of the richness of the data, the power-law relationship extends over 5 decades (from $10^3$ to $10^8$ ms).

The fitted refractory kernel (see  Fig.~\ref{fig:results}b1)  shows a strong reduction of touching rate during the first few hundreds of milliseconds after the last touch. For other subjects, it can even display a small increase in touching rate about 1s after the last touch (see  Fig.~\ref{fig:results}b2). This smooth transition from short ITI to longer ITI  removes the need to define an arbitrary onset of the power-law distribution \cite{Clauset09a}.

The fitted touch priority distribution (see  Fig.~\ref{fig:results}c1) (assuming that the \Other priority distribution is given by $q(y) = 1$) diverges for small priorities (which is the case when $a<1$). 
%
%
We repeated this fitting procedure for the 84 subjects. The population results are displayed on Fig.~\ref{fig:results}a2-c2. We found that over the population the priority parameter $a$ is fairly scattered around a median value of $a = 0.53$ (for model $M_6$) and of $a = 0.49$ (for model $M_5$). 
The large inter-individual differences is also highlighted in Fig.~\ref{fig:results}e which displays a broad distribution of touching rate $\rho$ over the population. 

\subsection{Model comparison}
To compare the different models (see Table~\ref{tab:models}) for each individual, we can use the Bayesian Information Criterion  (BIC) which is well suited for large data sets (i.e. large number of touching intervals $N$) which is precisely our case. BIC is given by $BIC = \log(N)|\theta|-2\mathcal{L}(\theta^*)$ where $|\theta|$ is the number of parameters and $\mathcal{L}(\theta^*)$ is the objective function given by Eq.~(\ref{eq:logpost}) and is evaluated at the MAP parameter $\theta^*$. To compare the different models for the whole population of $S=84$ subjects, we can define a population BIC in an analogously to the individual BIC given above. Let  $N_{\rm pop} = \sum_{s=1}^S N(s)$ denote the total number of inter-touch intervals of the whole population where $N(s)$ is the number of data points of subject $s$. Let $|\theta_{\rm pop}| = S|\theta|$ denote the total number of fitted parameters and let $\mathcal{L}_{\rm pop}(\theta_{\rm pop}^*) = \sum_{s=1}^S\mathcal{L}(\theta^*(s))$ denote the population objective function where $\theta^*(s)$ denotes the fitted parameters of subject $s$. The population BIC is therefore given by
\begin{eqnarray}
BIC_{\rm pop} &=&  \log(N_{\rm pop})S\abs{\theta}-2\sum_{s=1}^S\mathcal{L}(\theta^*(s)).
\end{eqnarray}

\begin{figure}[htbp]
\flushleft{\bf a} \\
\flushleft\includegraphics[width=0.4\textwidth]{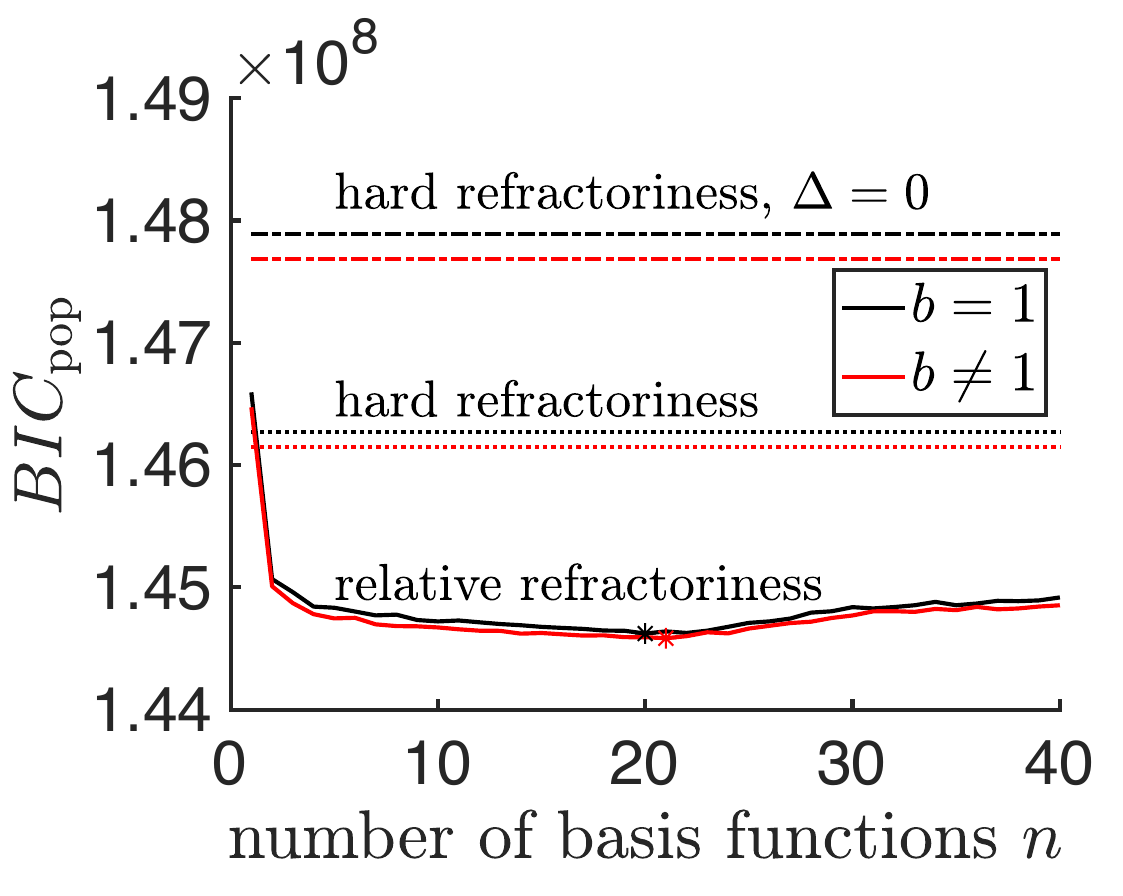}\\
\begin{center}
\begin{tabular}{ll}
{\bf b} & {\bf c}\\
\includegraphics[width=0.2\textwidth]{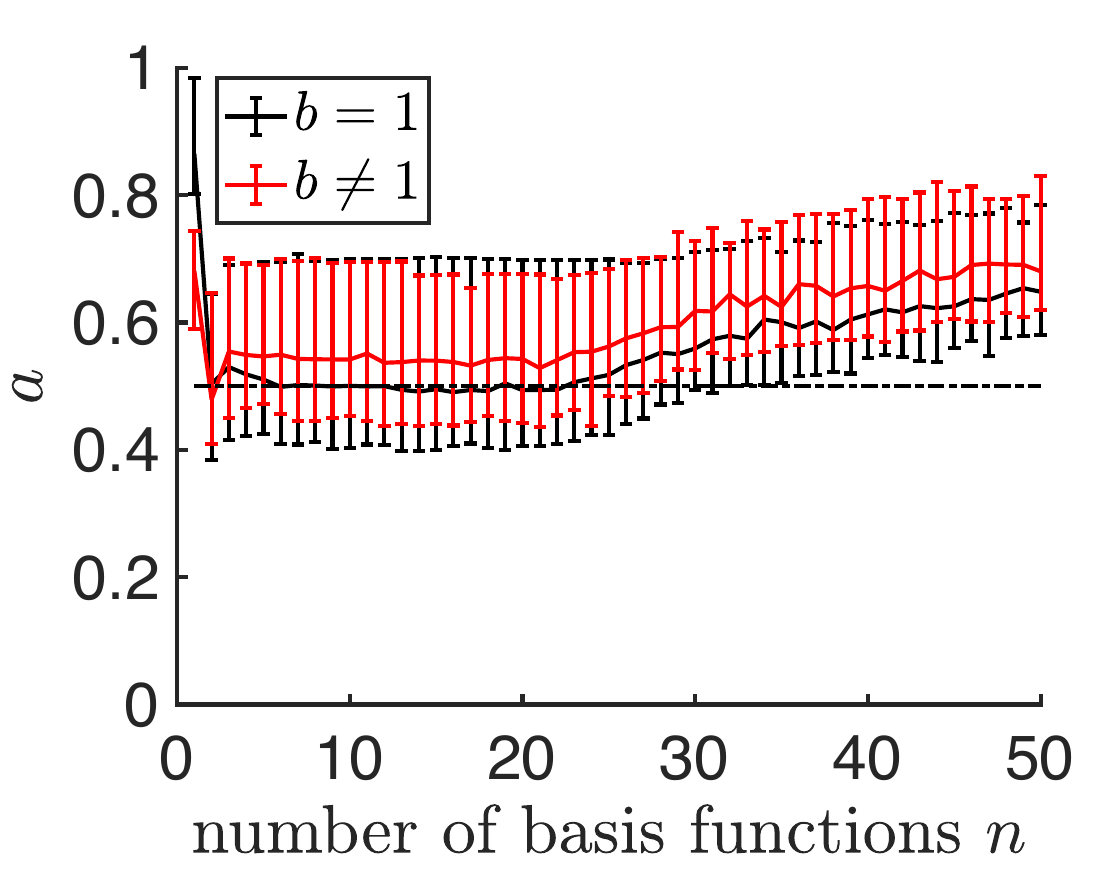}&
\includegraphics[width=0.2\textwidth]{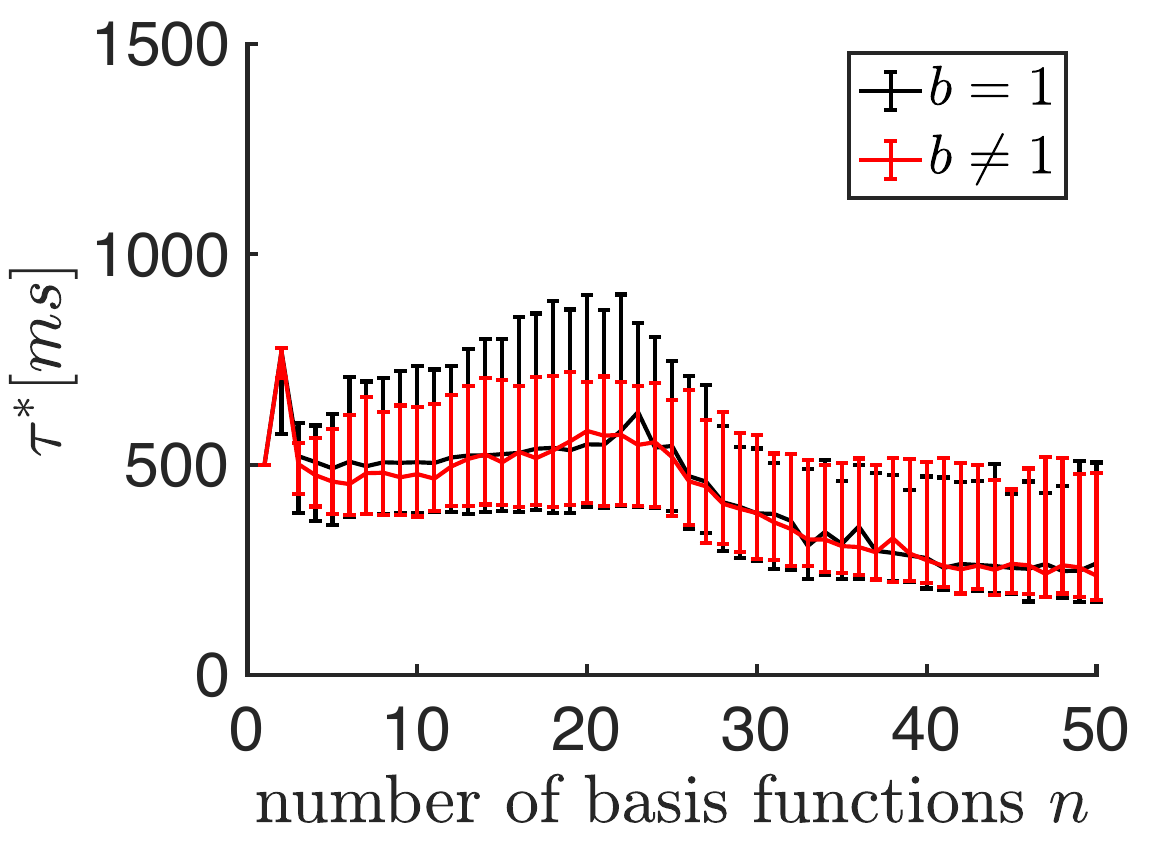}
\end{tabular}
\end{center}
\caption{Model comparison. {\bf a} Comparison of the normalized BIC for the 6 different models as a function of the number of basis functions $n$. Overall, models with relative refractoriness ($M_5$ and $M_6$, solid lines) outperform  models with hard refractoriness ($M_3$ and $M_4$, dotted lines; $M_1$ and $M_2$, dot-dashed lines). The best number of basis functions for model $M_5$ is $n^*=20$ (black star). The overall best model is $M_6$ with $n=21$ (red star). Note that the BIC difference between the red and the black star is  $-3.8\cdot 10^4$. {\bf b} Quantile plot for the priority parameter $a$ for model $M_5$ (black) and model $M_6$ (red). Solid lines denote the median across subjects and the error bars denote 25 and 75 percentiles. ({\bf c}) Same as {\bf b} but for the effective time constant $\tau^*$.}
\label{fig:s1}
\end{figure}

We found that the simplest models without any refractoriness ($M_1$ and $M_2$) or with hard refractoriness ($M_3$ and $M_4$) are outperformed by models with relative refractoriness ($M_5$ and $M_6$), see Fig.~\ref{fig:s1}. Indeed, despite their relative large number of parameters which penalizes the BIC, the models with relative refractoriness have a better (i.e. lower) BIC than the other models since they better describe short intervals. In particular, we found that the overall best model is $M_6$ with $n=21$ basis functions. When the priority parameter $b$ is set to one, then the best model of $M_5$ is when $n=20$. Note that the difference between the difference in BIC between the best model $M_6$ and the best model $M_5$ is $\Delta BIC_{\rm pop} = -3.8\cdot 10^4$ which is highly significant.

\subsection{Short vs long intervals}

\begin{figure}[htbp]
\begin{center}
\begin{tabular}{ll}
{\bf a} & {\bf b}\\
\includegraphics[width=0.2\textwidth]{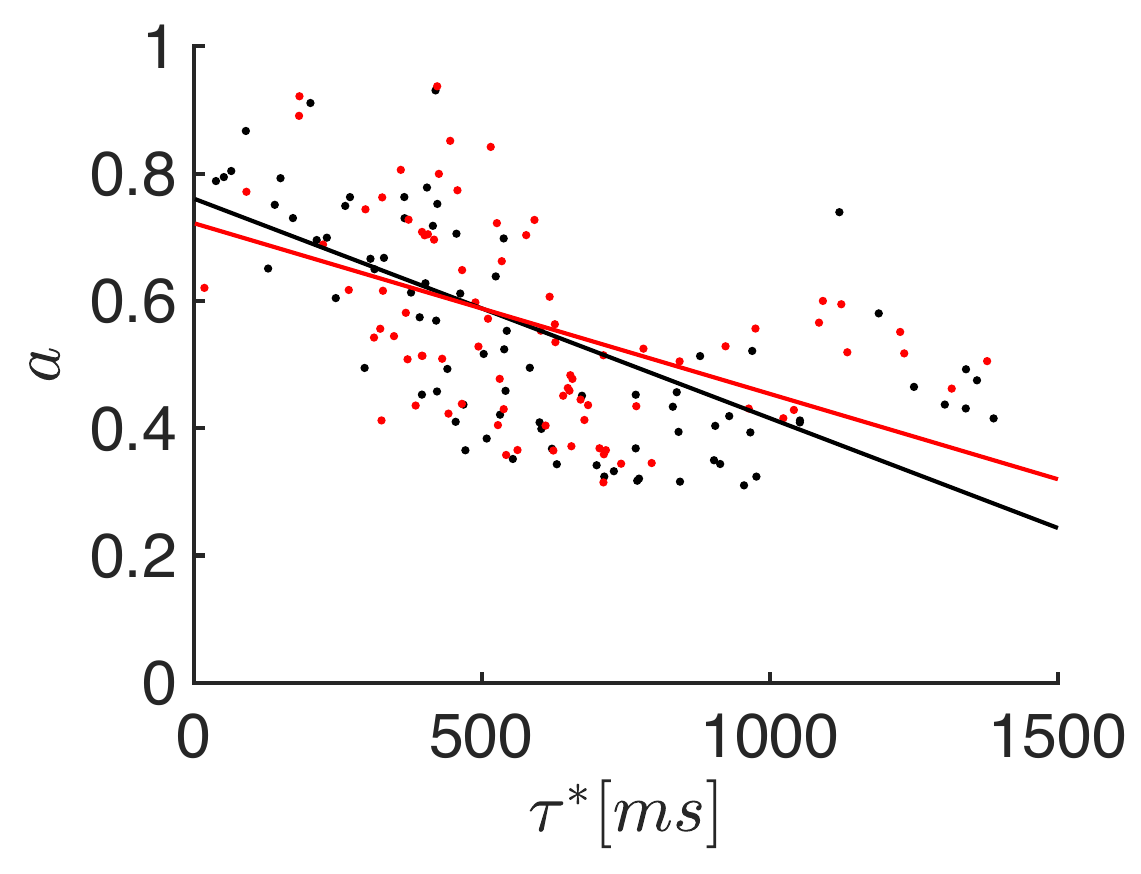}&
\includegraphics[width=0.2\textwidth]{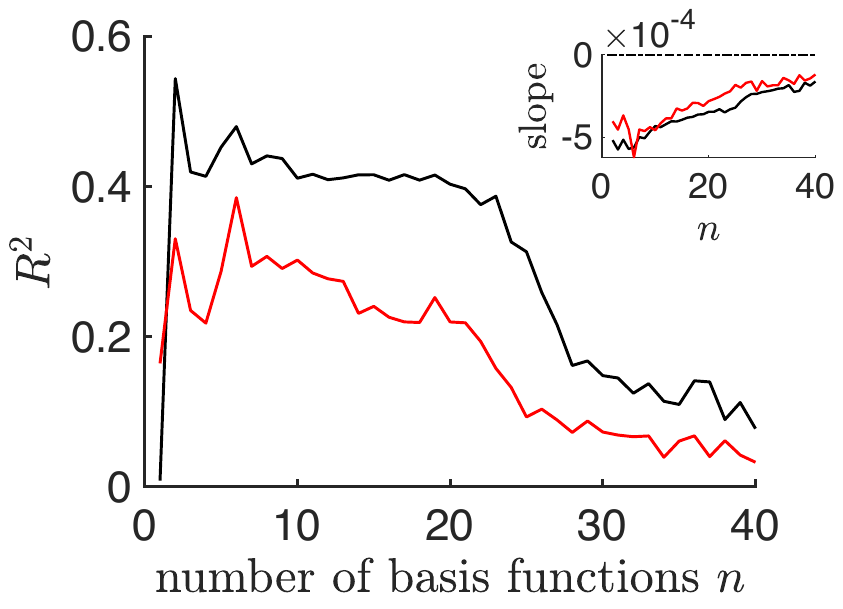}\\
{\bf c} & {\bf d}\\
\includegraphics[width=0.2\textwidth]{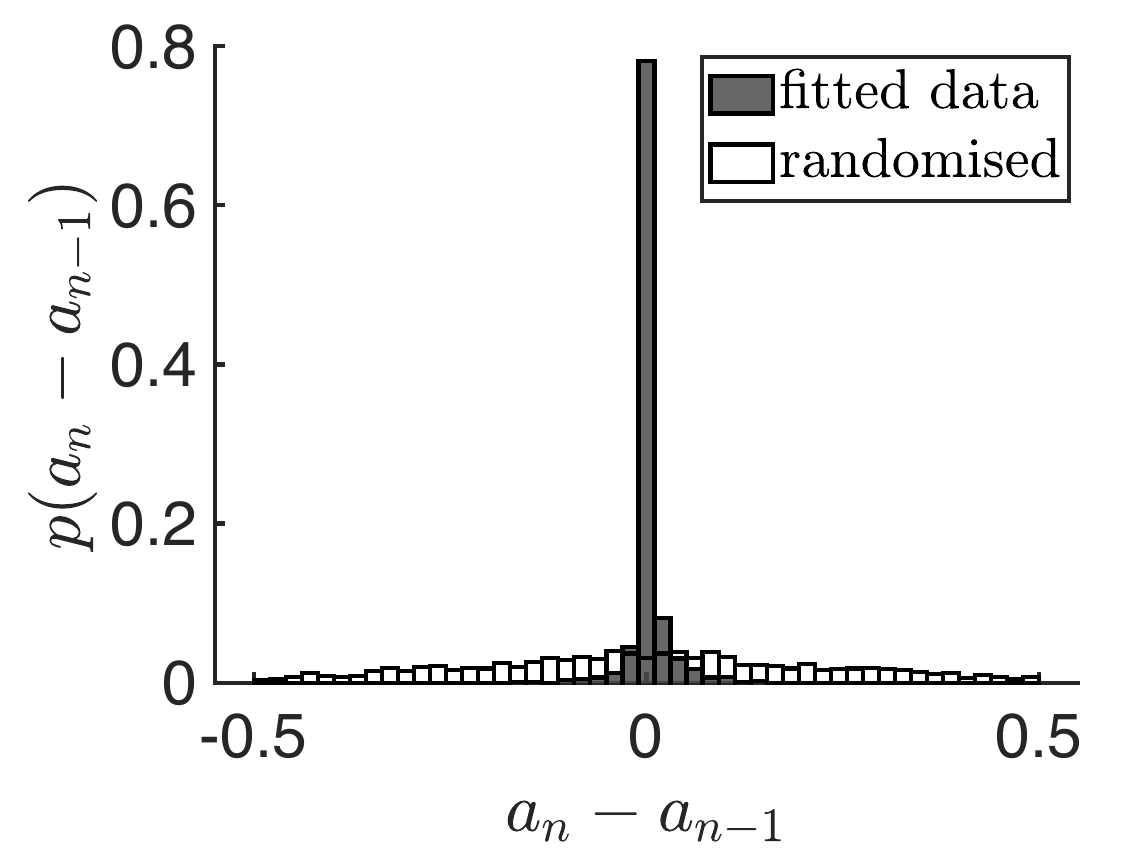}&
\includegraphics[width=0.2\textwidth]{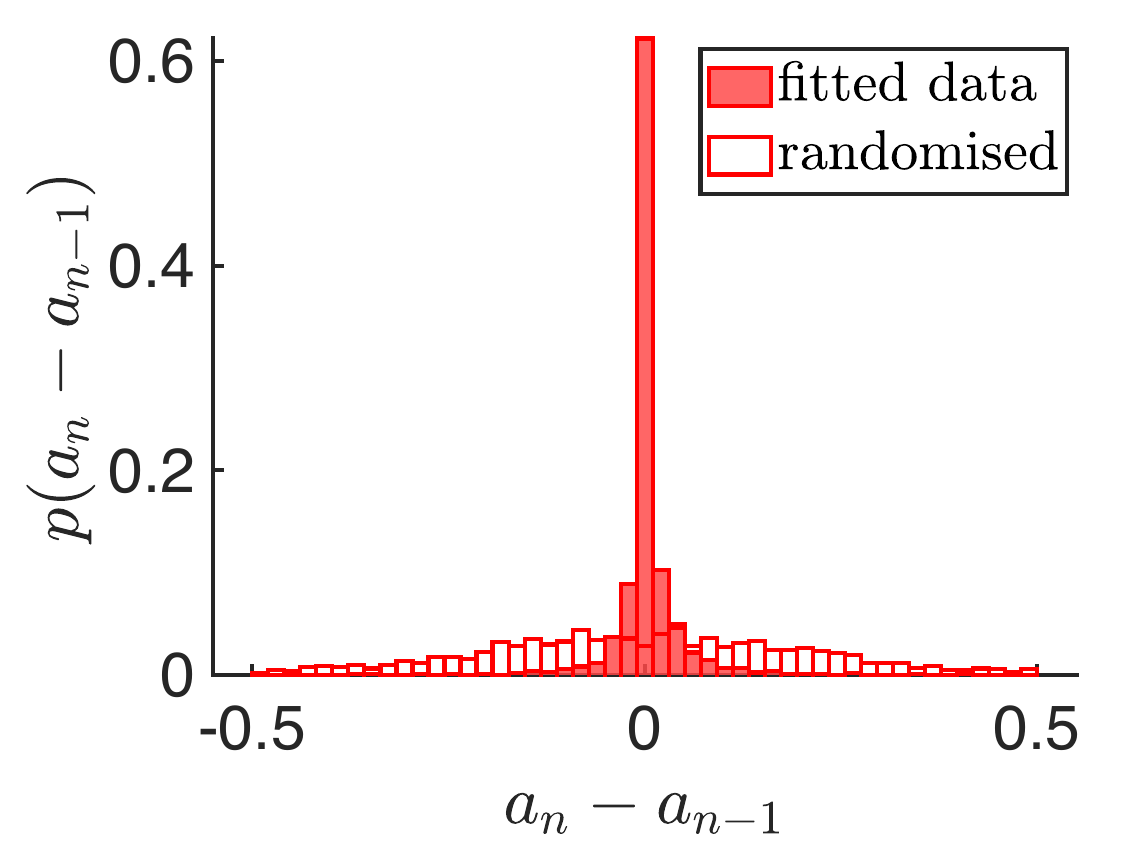}\\
{\bf e} & {\bf f}\\
\includegraphics[width=0.2\textwidth]{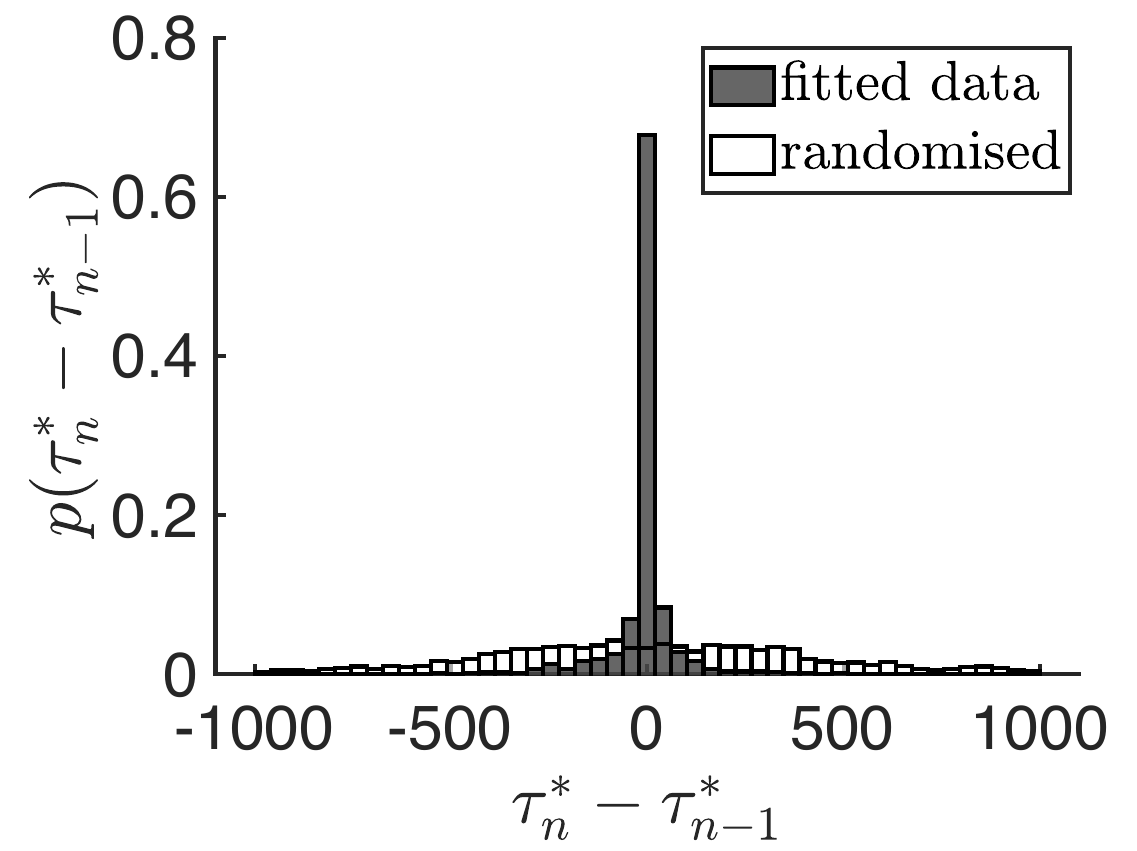}&
\includegraphics[width=0.2\textwidth]{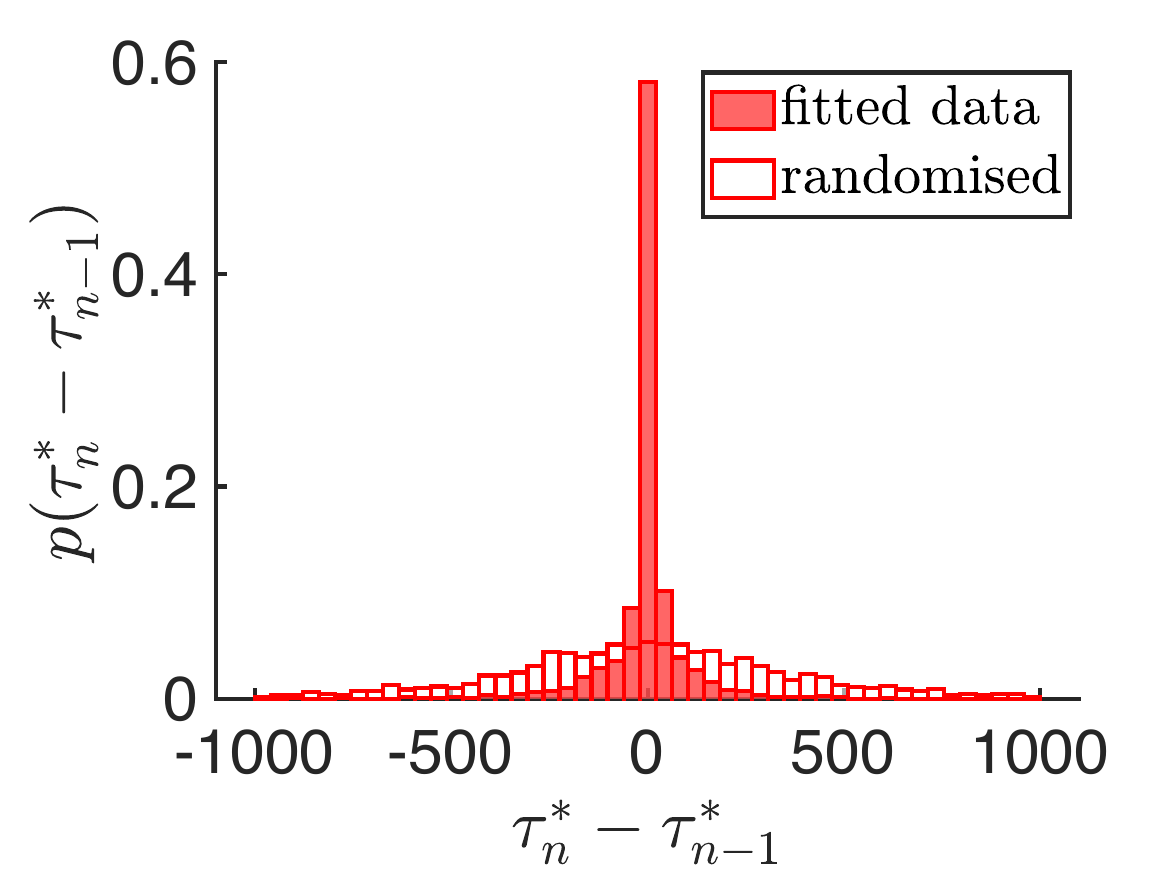}

\end{tabular}
\end{center}
\caption{{\bf a}. The scale-free exponent $a$ is inversely correlated with the effective refractory time constant for model $M_5$ ($b=1$, black) and $M_6$ ($b\neq 1$, red). {\bf b}. Explained variance ($R^2$) for this inverse correlation as a function of the number of basis functions. For the best model $M_5, (n=20)$, $R^2 = 40\%$ and for the best model $M_6, (n=21)$, $R^2 = 22\%$. For all numbers of basis function, the slope is negative (see {\bf inset}).  {\bf c}-{\bf f}: consistency of the fitted parameters. {\bf c}. The distribution of the difference $a_n(s)-a_{n-1}(s)$ across subjects $s=1,\dots,84$ and across basis function numbers $n=10,\dots,30$ (black, $\sigma_{\rm fit} = 0.022$) is much narrower than the distribution of the differences $a_n(s)-a_{n-1}(s')$ where the indices $s'$ are randomly permuted (white, $\sigma_{\rm rand} = 0.28$). {\bf d} Same as {\bf c}, but for model $M_6$ ($\sigma_{\rm fit} = 0.043$, $\sigma_{\rm rand} = 0.23$). {\bf e}. Similarly as in {\bf d}, the distribution of the difference of effective refractory time constants  $\tau^*_n(s)-\tau^*_{n-1}(s)$ (black, $\sigma_{\rm fit} = 88.7$ ms) much narrower than the one obtained when subjects are permuted for each $n$ (white,  $\sigma_{\rm rand} = 458$ ms). {\bf f}. Same as {\bf e} but for model $M_6$ ($\sigma_{\rm fit} = 100$ ms, $\sigma_{\rm rand} = 366$ ms).}
\label{fig:atau}
\end{figure}

Given the fairly broad distribution of fitted power-law exponent $a$ (Fig.~\ref{fig:s1}b) and effective time constant $\tau^*$ (Fig.~\ref{fig:s1}c), one could wonder whether this is a fitting artifact (which would come from a fairly flat landscape of the objective function for every subject) or whether the variability of those parameters actually comes from subject-to-subject variability. To test this, we compared the within-subject variability with the between-subject variability of those parameters and found a high degree of fitting consistency between different instantiations (i.e. different number of basis functions) of the same model (see Fig.~\ref{fig:atau}c-f). 

We then asked whether the effective refractory time constant $\tau^*$  is correlated with the power-law exponent across different subjects. Note that from the way the model is constructed, those two parameters are a priori unrelated since the refractory affects short inter-touch intervals and the power-law exponent affects longer intervals (see Fig.~\ref{fig:model}). We found that $a$ and $\tau^*$ are indeed inversely correlated (Fig.~\ref{fig:atau}a) with an explained variance of 40\% for model $M_5$ and 22\% for model $M_6$ (Fig.~\ref{fig:atau}b). This indicates that subjects that have a fast motor control (i.e. have a small $\tau^*$) also put a higher priority on their smartphone (i.e. higher $a$) in the sense that they have less low priority smartphone tasks.  

\section{Discussion}
%
We proposed a generalized priority-based model which is both flexible and tractable. The flexibility comes from the set of basis functions which describe refractory effects at short inter-touch intervals while the tractability stems from the simplified structure of the generative model in continuous-time which enables a fast fitting procedure. The flexibility is essential to capture inter-individual differences in touching behavior while the tractability is crucial for fitting large data sets. We found that that the inter-individual differences in low level motor control ability (reflected by the effective refractory time constant) can be partially explained by the higher-level cognitive processes which attributes priority to specific tasks (reflected by the priority parameter $a$). 

Other models describing smartphone activity have been proposed.  However they are aimed at addressing different questions (mostly sleep related) and use different type of data. The model of \cite{Cuttone17a} aims at predicting sleep patterns and rely on app launch timings binned over 15 minutes duration and therefore lack the possibility to describe refractory effects on the tens of milliseconds resolution. \cite{Abdullah14a} only considered screen-on and screen-off events to predict sleep patterns. 

More generally, circadian rhythms have received recently a lot of attention \cite{Aledavood15a,Aledavood15b,Aledavood18a} and the question has been addressed whether circadian rhythms could explain heavy tail distribution. \cite{Malmgren08a} has argued that a cascade of Poisson processes can give rise to power-law distribution. Interestingly, \cite{Jo12a} showed that even if the data is de-seasoned (i.e. the circadian and weekly patterns are removed from the time series of mobile phone events), the heavy-tails remain. 

Closer to the present study, the priority model proposed by \cite{Barabasi05a,Vazquez06a,Oliveira09a} already captures the power-law structure of the inter-event distribution. However, our model deviates significantly from their approach in several aspects. %
First, on the conceptual level, those authors stress the universality of the various behaviors. The claim is that certain activities such as browsing the Web, sending emails or loaning books fall into a specific universality class with power-law exponent of $\alpha = 1$ while other activities such as writing mails follow another universality class with exponent of $\alpha = 3/2$. Here, we found that the power-law exponent (averaged over the population) is $\alpha = a+1 \simeq 1.56 \pm 0.16$ which is indeed close to the rational exponent of $3/2$. However, it should be noted that the power-law exponent of individuals are fairly spread ranging from $\alpha = 1.31 \pm 0.04$ to $\alpha = 2.19 \pm 0.04$ which are clearly different from $\alpha = 1.5$. 

Capturing those non-universal exponents is possible in our model since the power-law exponent is given by $a+1$ where $a$ can take any real positive value. In contrast, in the work of \cite{Oliveira09a}, the exponent is determined by the length of the list of tasks\footnote{Technically, \cite{Oliveira09a} assume that an event occurs only if the interacting task for both agents A and B have the highest priority compared to all other tasks of length $L_A$ (for agent A) and $L_B$ for agent B. If $L = L_A = L_B$, then the power-law exponent is given by $1+ 1/(L-1)$}.  

The second difference w.r.t the studies of \cite{Barabasi05a,Vazquez06a,Oliveira09a} is that our model has been actually fitted to the whole set of event times (the touch times) such that we did not neglect small inter-event intervals by defining a (somewhat artificial) onset of the power-law distribution \cite{Clauset09a}. This is possible in our model since short intervals are captured by the refractory kernel. Note that even though refractory kernels have been used in other fields (e.g. in spiking neuron models, the probability of generating a spike just after a first one is also modulated by a refractory kernel \cite{Pillow08b,Gerstner14a,Surace15a}), the particularity here is that the specific form of the refractory kernel is such that its integral can be computed analytically which boosts the computational efficiency. 

Finally, fitting our model to the touching data has been possible because we considered the continuous-time priority model. Indeed, the marginal likelihood can be expressed analytically for the continuous-time model (and not for the discrete time model) which makes the maximum-likelihood parameter learning extremely efficient.

A separate line of research based on biological signals has also encountered scale-invariant relationships referred as $1/f$ pink noise \cite{Gilden95a,Chen97a}. However, those studies compute the power-spectrum density and not the inter-event distribution. Actually, if we do compute the power-spectrum density for the smartphone touching model, we found that in the limit of large frequencies, the power-spectrum density remains constant and does not decrease as $1/f$. 


Here, this generalized priority-based model has been applied to smartphone touching data, but could be applied to other event-based data sets which display power-law property for large inter-event intervals such as surface mails, emails or even foraging patterns. \\


\emph{Data accessibility}. The simulations have been performed using Matlab. The code and the smartphone touching data are available online \footnote{\protect{w}ww.ini.uzh.ch/$\sim$jpfister/code/Pfister\_18a\_code.zip}.\\

\emph{Author contribution}.  JPP and AG designed the study, developed the model used here and drafted the manuscript. JPP analyzed the data and formulated the mathematical model, aided by AG. AG collected the data and conceived the project with JPP.\\

\emph{Competing interests}. AG is an inventor of the patent-pending technology used to track touchscreen interactions in this study. AG and JPP are co-founders of QuantActions GmbH, a company focused on quantifying human behavior through smartphone interactions.\\

\emph{Acknowledgements}. We thanks Simone Carlo Surace for very helpful discussions. \\

\emph{Funding}. JPP was supported by the Swiss National Science Foundation grants PP00P3\_150637 and PP00P3\_179060. AG was supported by the Society in Science the Branco Weiss Fellowship and a research grant from the Holcim Foundation.

\appendix
\beginappendix

\section{}
\subsection{Invariance of the model}
\label{sec:inv} 
In this section, we will show that that the ITI distribution remains unchanged if the pair of priority distribution $(p(x),q(y))$ is replaced by $(\tilde{p}(x),\tilde{q}(y))$ where $\tilde{p}(x)$ and $\tilde{q}(y)$ are given by Eq.~(\ref{eq:ptilde}).

Let us consider the following change of variable: $x= \phi(x')$. The ITI distribution can be therefore expressed as
\begin{equation}
p(\tau) = \int_0^1 p^q(\tau|\phi(x'))p(\phi(x'))\phi'(x')dx',
\end{equation} 
where the conditional ITI distribution $p^q(\tau|\phi(x'))$ depends on the \Other priority distribution $q(y)$ via the instantaneous rate $\bar{\lambda}^q(\phi(x'),\tau)$ which can be expressed as 
\begin{eqnarray}
\bar{\lambda}^q(\phi(x'),\tau) &=& \rho r(\tau)\int_0^{\phi(x')}q(y)dy\\
&=& \rho r(\tau)\int_0^{x'}q(\phi(y))\phi'(y)dy = \bar{\lambda}^{\tilde{q}}(x',\tau),\nonumber
\end{eqnarray}
where $\tilde{q}$ is given by Eq.~(\ref{eq:ptilde}). Note that the dependence on $q$ is included only here for the clarity of the argument, but is omitted otherwise for the simplicity of the notation. Therefore the ITI distribution is invariant under the change of variable $\phi$ for both $x$ and $y$. Indeed, we have
\begin{eqnarray}
p(\tau) &=& \int_0^1 p^q(\tau|x)p(x)dx \nonumber \\
&=& \int_0^1 p^{\tilde{q}}(\tau|x)\tilde{p}(x)dx
\end{eqnarray}
For example, if the \touch priority distribution is given by $p(x) = {\rm Beta}(x;a,1)$ and the \Other priority distribution is given by $q(y) = {\rm Beta}(y;a',1)$, then the function $\phi(x)=x^k$ allows to generate a family of equivalent pairs of priority distributions $(\tilde{p}(x),\tilde{q}(y)) = ({\rm Beta}(x;ka,1),{\rm Beta}(y;ka',1))$. Therefore, the ITI remains unchanged as long as $a/a'$ remains constant. 

Note that this argument can be generalized to arbitrary smooth distribution $\tilde{p}$ and $\tilde{q}$. Let $\tilde{A}(x)$ denote the ratio of the logarithm of both cumulative density functions $\tilde{P}(x) = \int_0^x x'\tilde{p}(x')dx'$ and $\tilde{Q}(x) = \int_0^x x' \tilde{q}(x')dx'$:
\begin{equation}
\tilde{A}(x) = \frac{\log(\tilde{P}(x))}{\log(\tilde{Q}(x))}
\end{equation}
Since $\tilde{p}$ and $\tilde{q}$ are smooth, when $x\rightarrow 0$, we can express those priority distribution as $p(x)\simeq c_1x^{a-1}$ and $q(y) = c_2y^{a'-1}$. Also, since $\phi(x)$ is smooth, it can be approximated as $\phi(x) = x^k$ in the vicinity of $x=0$. Now we can show that the $\tilde{A}(0)$ is independent of the change of variable function $\phi$. Indeed
\begin{eqnarray}
\tilde{A}(0) &=& \lim_{x\rightarrow 0} \frac{\log(x\tilde{p}(x))}{\log(x\tilde{q}(x))} \nonumber \\
&=&  \lim_{x\rightarrow 0} \frac{\log(p(\phi(x))) + \log(\phi'(x)) + \log(x)}{\log(q(\phi(x))) + \log(\phi'(x)) + \log(x)} \nonumber\\
&=&  \lim_{x\rightarrow 0}\frac{\log(c_1) + ka\log(x)}{\log(c_2) + ka'\log(x)} = \frac{a}{a'}
\end{eqnarray}

is independent of $k$. 

\subsection{Log-likelihood gradient}

For the models with relative refractoriness, we fitted the parameters $\theta = (a,b,c,\gamma_1,\dots,\gamma_n)$ by performing maximum likelihood with a suitable regularization for the parameters $\gamma_i$.  Note that for a practical implementation, it is easier to learn $c = \log(\rho)$ instead of $\rho$ itself. For a set of inter-touch intervals $\mathcal{D} = \{\tau_i\}_{i=1}^N$, the log-likelihood can be expressed as
\begin{equation}
L(\theta) = Nc + \sum_{i=1}^N \log(r(\tau_i)) + \log\pa{\Ex{x E_i(x)}}, \label{eq:LL}
\end{equation}
where the expectation $\Ex{\cdot}$ is w.r.t $p(x) = {\rm Beta}(x;a,b)$ and the function $E_i(x)$ is given by
\begin{equation}
E_i(x) = \e^{-\rho x R(\tau_i)},
\end{equation}
and $R(\tau_i)$ is given by
\begin{equation}
R(\tau_i) := \int_0^{\tau_i}r(t)dt = \tau_i + \sum_{k=1}^n\frac{\gamma_k}{\alpha_k}\pa{1-\e^{-\alpha_k\tau_i}}.
\end{equation}
By noting that
\begin{equation}
\frac{\partial \log(p(x))}{\partial a} = \log(x) - \Ex{\log(x)},
\end{equation}
we can compute the log-likelihood gradient w.r.t $a$:
\begin{equation}
\frac{\partial L}{\partial a} = \sum_{i=1}^N\frac{{\rm cov}(xE_i(x),\log(x))}{\Ex{xE_i(x)}}.
\end{equation}
By symmetry, the gradient of $L$ w.r.t to $b$ yields
\begin{equation}
\frac{\partial L}{\partial b} = \sum_{i=1}^N\frac{{\rm cov}(xE_i(x),\log(1-x))}{\Ex{xE_i(x)}}.
\end{equation}
The gradient of $L$ w.r.t $c$ is given by
\begin{equation}
\frac{\partial L}{\partial c} = N-\rho\sum_{i=1}^N\frac{\Ex{x^2E_i(x)}}{\Ex{xE_i(x)}}R(\tau_i),
\end{equation}
Finally, the gradient of $L$ w.r.t $\gamma_k$ can be expressed as
\begin{eqnarray}
\frac{\partial L}{\partial \gamma_k} &=& \sum_{i=1}^N\frac{\partial r(\tau_i)/\partial \gamma_k}{r(\tau_i)}-\rho\frac{\Ex{x^2E_i(x)}}{\Ex{xE_i(x)}}\frac{\partial R(\tau_i)}{\partial \gamma_k}\nonumber\\
&=& \sum_{i=1}^N\frac{\e^{-\alpha_k\tau_i}}{r(\tau_i)} - \rho\frac{\Ex{x^2E_i(x)}}{\Ex{xE_i(x)}}\frac{\pa{1-\e^{-\alpha_k\tau_i}}}{\alpha_k}
\end{eqnarray}

For the models with hard refractoriness, the integral over the refractory kernel is given by
\begin{equation}
R(\tau_i) = \tau_i-\Delta \quad
\end{equation}
when $0\leq \Delta < \tau_{\rm min}$ where $ \tau_{\rm min} = \min_i\tau_i$ is the minimal inter-tap interval. Under this condition, the log-likelihood gradient yields
\begin{equation}
\frac{\partial L}{\partial \Delta} = \rho\frac{\Ex{x^2E_i(x)}}{\Ex{xE_i(x)}}\label{eq:dLDelta}
\end{equation}
which is positive as long as $\Delta\leq\tau_{\rm min}$.  When $\Delta >\tau_{\rm min}$, the log-likelihood goes to $-\infty$. Therefore the optimal refractory time constant is $\Delta^* = \tau_{\rm min}$.

\subsection{Computing the integrals}

Both the log-likelihood $L$ as well its gradient w.r.t to the parameters $\theta$ contain integrals that are delicate to evaluate. Indeed, the integrand of all those integrals depend on the Beta distribution ${\rm Beta}(x;a,b)$ which can diverge at $x=0$ or $x=1$ depending on the parameters $a$ and $b$. So whenever possible, we compute those integrals analytically. This can be done for the following integrals
\begin{equation}
\Ex{\log(x)}_{a,b} = \frac{d}{da}B(a,b) = \psi(a)-\psi(a+b),
\end{equation}
where $\psi(z) = d\log \Gamma(z)/dz$ is the digamma function and $B(a,b) = \Gamma(a)\Gamma(b)/\Gamma(a+b)$ is the Beta function. By symmetry, we have 
\begin{equation}
\Ex{\log(1-x)}_{a,b} = \psi(b)-\psi(a+b). \label{eq:log1x}
\end{equation}
By Taylor expanding the exponential in the expression of $E_i(x)$, the integral $\Ex{xE_i(x)}_{a,b}$ can be expressed as
\begin{eqnarray}
\Ex{xE_i(x)}_{a,b} &=& \frac{a}{a+b}\Ex{E_i(x)}_{a+1,b} \\
&=& \frac{a}{a+b}\F(a+1,a+b+1;-\rho R(\tau_i)) \nonumber
\end{eqnarray}
where $\F$ is the hypergeometric function defined as
\begin{equation}
\F(a,b;z) = \sum_{k=0}^{\infty}\frac{z^k}{k!}\frac{(a)_k}{(b)_k}
\end{equation}

and $(a)_k = \prod_{i=0}^{k-1}(a+k)$ for $k\geq 1$ (and $(a)_0=1$) is the rising factorial (also called Pochhammer function). Similarly, $\Ex{x^2E_i(x)}_{a,b}$ can be expressed as
\begin{equation}
\Ex{x^2E_i(x)}_{a,b} = \frac{B(a+2,b)}{B(a,b)}\F(a+2,a+b+2;-\rho R(\tau_i)).
\end{equation}
When it is not possible to compute the integrals analytically, the idea is to express the integral into a sum of two integrals where the first one is well suited for a numerical integration and the second one can be performed analytically. For example $\Ex{xE_i(x)\log(1-x)}_{a,b}$ can be computed as
\begin{eqnarray}
&&\Ex{xE_i(x)\log(1-x)}_{a,b} \nonumber \\
&=&\frac{a}{a+b}\left\{\Ex{\pa{E_i(x)-E_1(x)\log(1-x)}}_{a+1,b}\right.\nonumber\\
&+&\left. E_i(1)\Ex{\log(1-x)}_{a+1,b}\right\},
\end{eqnarray}
where the first term of the r.h.s can be computed numerically and the second term can be computed with Eq.~(\ref{eq:log1x}).

Finally, It should be noted that the integral $\Ex{xE_i(x)\log(x)}_{a,b} $ can be computed numerically straightforwardly since the integrand does not diverges when $x=0$ nor when $x=1$.

%

%

\section{}
\subsection{Smartphone data collection}
\label{sec:data}
A custom-designed software application (app, Touchometer) that could record the touchscreen events with a maximum error of 5 ms \cite{Balerna18a} was installed on each participant's phone. To determine this accuracy, controlled test touches were done at precisely 150, 300 and 600 ms while the Touchometer recorded at 147, 301 and 600 ms respectively, with standard deviations less than 15 ms (interquartile range less than 5 ms). The app posed as a service to gather the timestamps of touchscreen events that were generated when the screen was in an unlocked state. The operation was verified in a subset of phones by using visually monitored tactile events. The data were stored locally and transmitted by the user at the end of the study via secure email. One subject was eliminated as the app intermittently crashed after a software update.  The smartphone data were processed by using MATLAB (MathWorks, USA).

%
%


\bibliography{bibliography}

\end{document}